
\input harvmac.tex
\input epsf


\def\defn#1{\bigskip\noindent{\bf Definition #1} }
\def\thm#1{\bigskip\noindent{\bf Theorem #1} }
\def\lem#1{\bigskip\noindent{\bf Lemma #1} }
\def\prop#1{\bigskip\noindent{\bf Proposition #1} }

\def\rmk#1{\bigskip\noindent{\bf Remarks} }


\def\figin{\epsfcheck\figin}\def\figins{\epsfcheck\figins}
\def\epsfcheck{\ifx\epsfbox\UnDeFiNeD
\message{(NO epsf.tex, FIGURES WILL BE IGNORED)}
\gdef\figin##1{\vskip2in}\gdef\figins##1{\hskip.5in}
\else\message{(FIGURES WILL BE INCLUDED)}%
\gdef\figin##1{##1}\gdef\figins##1{##1}\fi}
\def\DefWarn#1{}
\def\figinsert{\goodbreak\midinsert}
\def\ifig#1#2#3{\DefWarn#1\xdef#1{fig.~\the\figno}
\writedef{#1\leftbracket fig.\noexpand~\the\figno}%
\figinsert\figin{\centerline{#3}}\medskip\centerline{\vbox{\baselineskip12pt
\advance\hsize by -1truein\noindent\footnotefont{\bf
Fig.~\the\figno:} #2}}
\bigskip\endinsert\global\advance\figno by1}



\def\unlockat{\catcode`\@=11}
\def\lockat{\catcode`\@=12}

\unlockat

\def\newsec#1{\global\advance\secno by1\message{(\the\secno. #1)}
\global\subsecno=0\global\subsubsecno=0\eqnres@t\noindent
{\bf\the\secno. #1} \writetoca{{\secsym}
{#1}}\par\nobreak\medskip\nobreak}
\global\newcount\subsecno \global\subsecno=0
\def\subsec#1{\global\advance\subsecno
by1\message{(\secsym\the\subsecno. #1)}
\ifnum\lastpenalty>9000\else\bigbreak\fi\global\subsubsecno=0
\noindent{\it\secsym\the\subsecno. #1} \writetoca{\string\quad
{\secsym\the\subsecno.} {#1}}
\par\nobreak\medskip\nobreak}
\global\newcount\subsubsecno \global\subsubsecno=0
\def\subsubsec#1{\global\advance\subsubsecno by1
\message{(\secsym\the\subsecno.\the\subsubsecno. #1)}
\ifnum\lastpenalty>9000\else\bigbreak\fi
\noindent\quad{\secsym\the\subsecno.\the\subsubsecno.}{#1}
\writetoca{\string\qquad{\secsym\the\subsecno.\the\subsubsecno.}{#1}}
\par\nobreak\medskip\nobreak}

\def\subsubseclab#1{\DefWarn#1\xdef
#1{\noexpand\hyperref{}{subsubsection}%
{\secsym\the\subsecno.\the\subsubsecno}%
{\secsym\the\subsecno.\the\subsubsecno}}%
\writedef{#1\leftbracket#1}\wrlabeL{#1=#1}}
\lockat

\def\IL{\relax{\rm I\kern-.18em L}}
\def\IH{\relax{\rm I\kern-.18em H}}
\def\IR{\relax{\rm I\kern-.18em R}}
\def\IC{\relax\hbox{$\inbar\kern-.3em{\rm C}$}}
\def\IZ{\relax\ifmmode\mathchoice
{\hbox{\cmss Z\kern-.4em Z}}{\hbox{\cmss Z\kern-.4em Z}}
{\lower.9pt\hbox{\cmsss Z\kern-.4em Z}} {\lower1.2pt\hbox{\cmsss
Z\kern-.4em Z}}\else{\cmss Z\kern-.4em Z}\fi}
\def\CM {{\cal M}}
\def\CN {{\cal N}}
\def\CR {{\cal R}}
\def\CD {{\cal D}}
\def\CF {{\cal F}}

\def\CO {{\cal O}}
\def\CZ {{\cal Z}}

\def\CH {{\cal H}}
\def\CC {{\cal C}}
\def\CB {{\cal B}}
\def\CS {{\cal S}}

\def\CM {{\cal M}}
\def\CN {{\cal N}}

\def\CO {{\cal O}}

\def\CZ {{\cal Z }}
\def\CS {{\cal S }}

\font\manual=manfnt \def\dbend{\lower3.5pt\hbox{\manual\char127}}

\def\IZ{\relax\ifmmode\mathchoice
{\hbox{\cmss Z\kern-.4em Z}}{\hbox{\cmss Z\kern-.4em Z}}
{\lower.9pt\hbox{\cmsss Z\kern-.4em Z}} {\lower1.2pt\hbox{\cmsss
Z\kern-.4em Z}}\else{\cmss Z\kern-.4em Z}\fi}
\def\half {{1\over 2}}

\def\CM {{\cal M}}
\def\CN {{\cal N}}

\def\CO {{\cal O}}

\def\CZ {{\cal Z }}
\def\CS {{\cal S }}

\def\f#1#2{{#1 \over #2}}


\def\IZ{\relax\ifmmode\mathchoice
{\hbox{\cmss Z\kern-.4em Z}}{\hbox{\cmss Z\kern-.4em Z}}
{\lower.9pt\hbox{\cmsss Z\kern-.4em Z}} {\lower1.2pt\hbox{\cmsss
Z\kern-.4em Z}}\else{\cmss Z\kern-.4em Z}\fi}
\def\IB{\relax{\rm I\kern-.18em B}}
\def\IC{{\relax\hbox{$\inbar\kern-.3em{\rm C}$}}}
\def\ID{\relax{\rm I\kern-.18em D}}
\def\IE{\relax{\rm I\kern-.18em E}}
\def\IF{\relax{\rm I\kern-.18em F}}
\def\IG{\relax\hbox{$\inbar\kern-.3em{\rm G}$}}
\def\IGa{\relax\hbox{${\rm I}\kern-.18em\Gamma$}}
\def\IH{\relax{\rm I\kern-.18em H}}
\def\II{\relax{\rm I\kern-.18em I}}
\def\IK{\relax{\rm I\kern-.18em K}}
\def\IP{\relax{\rm I\kern-.18em P}}
\def\IQ{\relax\hbox{$\inbar\kern-.3em{\rm Q}$}}

\def\inbar{\,\vrule height1.5ex width.4pt depth0pt}

\def\mod{{\rm mod}}

\font\cmss=cmss10 \font\cmsss=cmss10 at 7pt
\def\IR{\relax{\rm I\kern-.18em R}}


\def\makeblankbox#1#2{\hbox{\lower\dp0\vbox{\hidehrule{#1}{#2}%
   \kern -#1
   \hbox to \wd0{\hidevrule{#1}{#2}%
      \raise\ht0\vbox to #1{}
      \lower\dp0\vtop to #1{}
      \hfil\hidevrule{#2}{#1}}%
   \kern-#1\hidehrule{#2}{#1}}}%
}%
\def\hidehrule#1#2{\kern-#1\hrule height#1 depth#2 \kern-#2}%
\def\hidevrule#1#2{\kern-#1{\dimen0=#1\advance\dimen0 by #2\vrule
    width\dimen0}\kern-#2}%
\def\openbox{\ht0=1.2mm \dp0=1.2mm \wd0=2.4mm  \raise 2.75pt
\makeblankbox {.25pt} {.25pt}  }
\def\opensquare{\ht0=3.4mm \dp0=3.4mm \wd0=6.8mm  \raise 2.7pt \makeblankbox
{.25pt} {.25pt}  }

\def\sector#1#2{\ {\scriptstyle #1}\hskip 1mm
\mathop{\opensquare}\limits_{\lower
1mm\hbox{$\scriptstyle#2$}}\hskip 1mm}

\def\tsector#1#2{\ {\scriptstyle #1}\hskip 1mm
\mathop{\opensquare}\limits_{\lower
1mm\hbox{$\scriptstyle#2$}}^\sim\hskip 1mm}


\def\inbar{\,\vrule height1.5ex width.4pt depth0pt}

\font\cmss=cmss10 \font\cmsss=cmss10 at 7pt
\def\IR{\relax{\rm I\kern-.18em R}}




\lref\bateman{Bateman Manuscript project}

\lref\bost{L. Alvarez-Gaum\'e, J.-B. Bost, G. Moore, P. Nelson,
and C. Vafa, ``Bosonization on higher genus Riemann surfaces,''
Commun.Math.Phys.112:503,1987 }

\lref\cardi{ K. Behrndt, G. Lopes Cardoso, B. de Wit, R. Kallosh,
D. L\"ust, T. Mohaupt, ``Classical and quantum N=2 supersymmetric
black holes,'' hep-th/9610105}

\lref\bloch{S. Bloch, ``The proof of the Mordell conjecture,''
Math. Intell. {\bf 6}(1984) 41}

\lref\cassels{J.W.S. Cassels, {\it Lectures on Elliptic Curves},
Cambridge University Press, 1995}

\lref\cox{D.A. Cox, {\it Primes of the form $x^2 + n y^2$}, John
Wiley, 1989.}

\lref\cremona{J.E. Cremona, {\it Algorithms for modular elliptic
curves} Cambridge University Press 1992} \lref\davenport{H.
Davenport, {\it Multiplicative Number Theory}, Second edition,
Springer-Verlag, GTM 74}

\lref\dewit{Bernard de Wit, Gabriel Lopes Cardoso, Dieter L\"ust,
 Thomas Mohaupt, Soo-Jong Rey,
``Higher-Order Gravitational Couplings and Modular Forms in
$N=2,D=4$ Heterotic String Compactifications,'' hep-th/9607184;
G.L. Cardoso, B. de Wit, and T. Mohaupt, ``Corrections to
macroscopic supersymmetric black hole entropy,'' hep-th/9812082}

\lref\dvv{R. Dijkgraaf, E. Verlinde, and H. Verlinde, ``Counting
Dyons in N=4 String Theory,'' hep-th/9607026;Nucl.Phys. B484
(1997) 543-561}

\lref\duke{W. Duke, ``Hyperbolic distribution problems and
half-integral weight Maass forms,'' Invent. Math. {\bf 92} (1988)
73} \lref\ez{Eicher and Zagier} \lref\faltings{G. Faltings,
``Finiteness theorems for abelian varieties over number fields,''
in {\it Arithmetic Geometry}, G. Cornell and J.H. Silverman, eds.
Springer 1986} \lref\fks{S. Ferrara, R. Kallosh, and A.
Strominger, ``N=2 Extremal Black Holes,''   hep-th/9508072}
\lref\fk{S. Ferrara and R. Kallosh, ``Universality of
Sypersymmetric Attractors,''   hep-th/9603090;  ``Supersymmetry
and Attractors,'' hep-th/9602136; S. Ferrara, ``Bertotti-Robinson
Geometry and Supersymmetry,'' hep-th/9701163} \lref\enneightbh{ L.
Andrianopoli, R. D'Auria, S. Ferrara, P. Fre', M. Trigiante, ``
$E_{7(7)}$ Duality, BPS Black-Hole Evolution and Fixed Scalars,''
 hep-th/9707087; Nucl.Phys. B509 (1998) 463-518
 }

\lref\fm{S. Ferrara and J. Maldacena, ``Branes, central charges
and $U$-duality invariant BPS conditions,'' hep-th/9706097}

\lref\husemoller{D. Husemoller, {\it Elliptic Curves},
Springer-Verlag, GTM 111}

\lref\faltings{G. Faltings, ``Calculus on arithmetic surfaces,''
Ann. Math. {\bf 119}(1984) 387}

\lref\stark{A. Granville and H. Stark, ``$ABC$ implies no `Siegel
zeroes','' preprint.}

\lref\bgross{B. Gross, {\it Arithmetic on elliptic curves with
complex multiplication}, Springer-Verlag LNM  776}

\lref\gzsm{B. Gross and D. Zagier, ``On singular moduli,'' J.
reine angew. Math. {\bf 355} (1985) 191}

\lref\kalkol{R. Kallosh and B. Kol, ``E(7) Symmetric Area of the
Black Hole Horizon,'' hep-th/9602014}

\lref\kawai{T. Kawai, ``K3 surfaces, Igusa cusp form and string
theory,'' hep-th/9710016} \lref\tkawai{T. Kawai, ``K3 surfaces,
Igusa cusp form and string theory,'' hep-th/9710016} \lref\msw{J.
Maldacena, A. Strominger, and E. Witten, ``Black Hole Entropy in
M-Theory,''  hep-th/9711053}

\lref\arthatt{G. Moore, ``Arithmetic and Attractors,''
hep-th/9807087 ; ``Attractors and Arithmetic,'' hep-th/9807056}

\lref\fareytail{ ``A Black Hole Farey Tail,'' unpublished. To
appear, some day, somewhere, maybe,....}

\lref\lang{S. Lang, {\it Introduction to Arakelov Geometry},
Springer-Verlag 1988}

\lref\langalg{S. Lang, {\it Algebra}, Addison-Wesley 1970}

\lref\langelliptic{S. Lang, {\it Elliptic Functions},
Addison-Wesley, 1973}

\lref\littlewood{J.E. Littlewood, ``On the class number of the
corpus $P(\sqrt{-k})$,'' Proc. Lond. Math. Soc., Ser. 2, {\bf 27},
Part 5. (1928)}

\lref\mazur{B. Mazur, ``Arithmetic on curves,'' Bull. Amer. Math.
Soc. {\bf 14}(1986) 207}

\lref\miller{S.D. Miller, ``Large Values of $\f{L'}{L}(1,-p)''$,
appendix to this paper.}

\lref\mumford{D. Mumford, Tata Lectures on Theta}

\lref\peet{A. Peet, ``The Bekenstein Formula and String Theory
(N-brane Theory),'' hep-th/9712253}

\lref\rohrlich{D. Rohrlich, ``Elliptic curves with good reduction
everywhere,'' J. London Math. Soc., {\bf 25}(1982)216}

\lref\silvermanag{J. Silverman, {\it Arithmetic Geometry}, G.
Cornell and J.H. Silverman, eds. Springer Verlag 1986}

\lref\silveraec{J. Silverman, {\it The Arithmetic of Elliptic
Curves}, Springer Verlag GTM 106 1986}

\lref\silveradvtop{J. Silverman, {\it Advanced Topics in the
Arithmetic of Elliptic Curves} Springer Verlag GTM 151, 1994}

\lref\stromi{A. Strominger, ``Macroscopic Entropy of $N=2$
Extremal Black Holes,''  hep-th/9602111}

\lref\sv{A. Strominger and C. Vafa, ``Microscopic Origin of the
Bekenstein-Hawking Entropy,'' hep-th/9601029; Phys.Lett. B379
(1996) 99-104 }

\lref\tatuzawa{T. Tatuzawa, ``On a theorem of Siegel,'' Jap. J.
Math. {\bf 21}(1951)163 }

\lref\weil{A. Weil, {\it Elliptic functions according to
Eisenstein and Kronecker}, Springer-Verlag 1976}

\lref\wittenls{E. Witten, `` Phases of $N=2$ Theories In Two
Dimensions,'' hep-th/9301042;Nucl.Phys. B403 (1993) 159-222}

\Title{ \vbox{\baselineskip12pt\hbox{hep-th/9903267}
 \hbox{YCTP-P8-99 }
\hbox{IASSNS-HEP-99/34   }}} {\vbox{\centerline{Landau-Siegel
Zeroes and Black Hole Entropy  }
\bigskip
\centerline{  }}}
\smallskip
\centerline{Stephen D.  Miller$^1$ and Gregory Moore$^{2,3}$ }
\medskip
\centerline{ $^1$ Department Of Mathematics, Yale University}
\centerline{ New Haven, Connecticut 06520, USA}\bigskip
\centerline{ $^2$ Department Of Physics, Yale University}
\centerline{ New Haven, Connecticut 06520, USA}
\medskip
\centerline{$^3$ School of Natural Sciences} \centerline{Institute
for Advanced Study} \centerline{Princeton, NJ 08540}
\bigskip
\baselineskip 18pt

\medskip
\noindent There has been some speculation about relations of
D-brane models of black holes to arithmetic. In this note we point
out that some of these speculations have implications for a circle
of questions related to the generalized Riemann hypothesis on the
zeroes of Dirichlet $L$-functions.

\noindent
\Date{April 1, 1999; Revised June 18, 1999; December 6, 1999}

\newsec{Introduction}

In \arthatt\  some connections were made between string theoretic
models of black holes and certain issues in arithmetic. One of the
more speculative suggestions in \arthatt\  was a proposal that the
entropy of BPS black holes is related to an arithmetic height of
certain arithmetic varieties. In the present note we remark on
some connections between these speculations and some issues
related to the Riemann hypothesis. In particular, the speculations
seem most relevant to the question of ``Landau-Siegel zeroes,''
which are hypothetical zeroes of $L$-functions very close to
$s=1$. (A precise definition is given in definition 4.4.)

In section 2 we review the theory of Strominger-Vafa \sv.  In
section 3 we summarize, reformulate, and sharpen
 the statements from \arthatt\ whose implications
we wish to explore.  In section 4 we provide some background
information on analytic number theory
 and Dirichlet L-functions, and in section 5 we see how
everything fits together.  In particular, we discuss a close
interplay between the Strominger-Vafa prediction about black hole
entropy, the Landau-Siegel zero, and the minimal discriminant of
an  elliptic curve with complex multiplication.

{\bf Warnings}: This note is written for a mixed audience of both
string theorists and analytic number theorists, so some trivial
things are explained. We caution the reader at the outset that the
evidence for the height conjecture of \arthatt\ is slim, to say
the least. Thus, this paper should be regarded as an exercise in
Pascal's Wager.

\newsec{Summary of the attractor mechanism and
the Strominger-Vafa proposal}

Let $X$ be a compact Calabi-Yau 3-fold, and let $\gamma\in
H_3(X;\IZ)$ be an integral homology class. String theory
associates two interesting mathematical objects to the pair
$(X,\gamma)$:

a.) A finite-dimensional Hermitian vector space $\CH(\gamma)$.

b.) Another   Calabi-Yau $X_\gamma$, in the same complex structure
moduli space as $X$.

We can interpret (a) and  (b) both mathematically and physically.
The physical setting is the theory of BPS black holes in
$d=4,\CN=2$ compactifications of type II string on a Calabi-Yau
manifold $X$. We now explain $(a)$ and $(b)$ in a little more
detail.

First consider $(a)$. In physics $\CH(\gamma)$ is the space of BPS
states of charge $\gamma$. The definition of this space has not
been completely rigorously formulated mathematically, although
this should be possible using the theory of $D$-branes. Very
roughly speaking $\CH(\gamma)$ should be defined mathematically as
follows. Consider the moduli space $\CM(\gamma)$ of pairs
$(\Sigma, A)$ where $\Sigma$ is a smooth special Lagrangian
submanifold of $X$ in the homology class $\gamma$, and $A$ is a
flat $U(1)$ connection on $\Sigma$. The moduli space $\CM(\gamma)$
inherits a metric from the Calabi-Yau metric on $X$, and
$\CH(\gamma)$ is   the $L^2$-cohomology of $\CM(\gamma)$. \foot{In
the mirror formulation $\gamma$ would specify the Chern classes of
a coherent sheaf in the mirror Calabi-Yau $\tilde X$, and
$\CM(\gamma)$ would be the $L^2$ cohomology of the moduli space of
such sheaves.}

Now consider  (b). By Yau's theorem, a Calabi-Yau manifold may be
specified by its complex structure and its K\"ahler class.  $X$
will belong to a family of Calabi-Yau manifolds with complex
structures in $\CM_{cplx}$.  (For what follows we will need to
work on the universal cover $\widetilde{\CM}_{cplx} $.) The map
$(b)$ is provided by the  ``attractor mechanism'' of Ferrara,
Kallosh, and Strominger \fks\stromi\fk. For each $\gamma$ there is
a dynamical system on the moduli space $\widetilde{\CM}_{cplx} $
and, for suitable $\gamma$, \foot{Some discussion of which
$\gamma$ are suitable is given in \arthatt.} the dynamical system
will have a unique fixed point $X_\gamma$. Associated with the
Calabi-Yau $X_\gamma$ is the normalized period \eqn\normlper{
\vert \CZ(\gamma) \vert^2 := {\vert \int_\gamma \Omega_\gamma
\vert^2 \over \vert \int_{X_\gamma} \Omega_\gamma \wedge \bar
\Omega_\gamma \vert } } where $\Omega_\gamma$ is a nowhere
vanishing holomorphic $(3,0)$ form on $X_\gamma$.

The connection between objects (a) and (b) is provided by  the
Strominger-Vafa proposal for the microstates accounting of black
hole entropy \sv. The idea may be summarized in the following four
steps:
\item{1.}
Given a Calabi-Yau $X$, one may write a system of partial
differential equations for a Minkowski-signature 4-manifold $M$
equipped with certain geometrical data (e.g. a rank $\half b_3(X)$
torus bundle with connection). These equations generalize the
Einstein equations on $M$ and are called the supergravity
equations. A choice of charge vector $\gamma$ is equivalent to a
choice of boundary conditions for these equations. For appropriate
vectors $\gamma$ (those for which the normalized period has an
isolated minimum in $\widetilde{\CM}_{cplx} $), the supergravity
equations admit black hole solutions $\CB(\gamma)$.  Using the
supergravity equations one may compute the   horizon area
$A(\gamma)$ of the black hole solution
 $\CB(\gamma)$.  It turns out that
\eqn\horizon{ A(\gamma) = 4 \pi \vert \CZ(\gamma) \vert^2. } We
work in Planck units $\ell_{planck} = 1$.

\item{2.} According to Bekenstein and Hawking
a black hole is a thermodynamical object. It has a temperature and
an entropy, and the latter is given by: \eqn\entrper{ S_{\rm
sugra}(\gamma) = { A(\gamma) \over  4}. }

\item{3.} According to  Boltzmann and Planck,
we have the exact formula: \eqn\boltz{ S_{\rm micro}(\gamma) =  k
\log[ W(\gamma) ] } where $W(\gamma)$ is the dimension of the
space of available states in the microcanonical ensemble specified
by the charges $\gamma$ and the energy $M= \vert \CZ(\gamma)
\vert$. Here $k$ is Boltzmann's constant; we henceforth choose
units with $k=1$.

\item{4.}
According to Strominger and Vafa, for charge vectors $\gamma$
which are in some sense large,  physical states formed by BPS
configurations of D3-branes in $X$ with charge $\gamma$ can be
described  macroscopically, in a supergravity approximation, by
the black hole solutions $\CB(\gamma)$ to the supergravity
equations.
 From the microscopic,
D-brane viewpoint we identify $W= \dim \CH(\gamma) $, for all
$\gamma$, large or small. The correspondence between
configurations of $D$-branes and the black hole solutions
  is formulated mathematically
as the statement that for large $\gamma$ \eqn\svprop{
 S_{\rm micro}(\gamma)\sim S_{\rm sugra}(\gamma).
}

Putting all the above statements together one arrives at   the
mathematically startling proposal that: \eqn\largegamm{
 \log [\dim \CH(\gamma) ]
\sim   \pi \vert \CZ(\gamma) \vert^2 } for large $\gamma$. Both
sides of   equation \largegamm\ are susceptible of precise
mathematical definition. Moreover,  such a    statement would be a
deep and surprising mathematical fact. Before that becomes a
reality, several clarifications of the meaning of \largegamm\ must
be carried out. In particular we need:

\item{1.} A precise definition of $\CH(\gamma)$
and a definition of  $\dim \CH(\gamma)$ (since $\CH(\gamma)$ is a
graded vector space, we might well have to use a graded
dimension.)

\item{2.}  A precise statement of the meaning of
``large $\gamma$.''

\item{3.}  A precise statement of the meaning of
the asymptotic symbol $\sim$.

It is quite necessary to take a limit of ``large $\gamma$,''  in
order to justify  the supergravity approximation to string theory,
and thereby the connection to black holes. We will take it to mean
operationally that we consider sequences $\gamma_n$ of charges
such that $z_n:=\vert \CZ(\gamma_n) \vert \rightarrow \infty$. We
refer to such a sequence of charges as a {\it big sequence}.

Another subtlety which we would like to mention is the concept of
{\it effective} asymptotics.  For example, one could make a
statement that a certain quantity is bounded, but without being
able to determine anything about the nature of the bound or even
being able to begin computing it. We will assume that the limiting
behaviors of the physical quantities are effective: that is, any
time a constant (implicit or named) is asserted to exist, one
could furthermore explicitly compute or name such a constant.

\subsec{Versions of the Strominger-Vafa conjecture}

Point 3 above (below \largegamm) has not really been addressed in
the literature. The Strominger-Vafa conjecture \largegamm\ in fact
admits several  inequivalent formulations. Two plausible versions
of the conjecture are:

\bigskip

\noindent{\bf SSV Conjecture.} ({\it Strong SV conjecture}): If
$\{ \gamma_n \}$ is a big sequence then $S_{\rm micro}(\gamma)$
has an asymptotic expansion in $z_n := \vert \CZ(\gamma_n) \vert$.
More precisely, \eqn\strngfrm{ {S_{\rm micro}(\gamma_n)  \over \pi
z_n^2} = 1 + \CO\biggl( {\log z_n \over  z_n^2} , {1 \over
z_n^\delta  } \biggr), } where we use $\CO$-notation in the
precise sense of asymptotics of sequences in $n$, and $\delta$ (a
quantity discussed below) is positive.

\bigskip

This is an expectation based on the supergravity approach to black
hole entropy for CY 3-fold black holes. See, for examples, the
discussions in \cardi\msw\dewit. The corrections arise from both
$M$-theory corrections to the 11D supergravity and from quantum
effects associated with the compactification. The quantity
$\delta$ in \strngfrm\ depends on details of the sequence
$\gamma_n$, e.g., on the details of the  directions of the
$\gamma_n$ in $H^3(X;\IZ)$ as $z_n$ goes to infinity. The SSV
conjecture might  appear to be obvious from the point of view of
supergravity. However, a systematic scheme for calculating
corrections to the leading order supergravity approximation to
$M$-theory is unkown. Moreover, it is known that supergravity,
which treats
 charges as continuous, can miss subtle
arithmetic properties. (For example, the number of $U$-duality
classes of $\gamma$'s with a fixed value of $\vert
\CZ(\gamma)\vert$ can be $1$ in supergravity, but in fact is given
by class numbers in the exact formulation  \arthatt.) The SSV
conjecture is also suggested by the proposed formulae for exact
results on $S_{\rm micro}(\gamma) $ in the case when $X = S \times
E$ with $S$ a K3 surface and $E$ an elliptic curve \dvv.  Using
the Cardy formula \foot{The Cardy formula is a generalization of
the Hardy-Ramanujan formula for the asymptotics of the partition
function and gives the asymptotics of conformal field theory
partition functions.  } one can justify the SSV conjecture
 for this special class of Calabi-Yau
3-folds. However, the existing proposals, based on elliptic genera
of symmetric products are only firmly justified for
 5D black holes, and the extension to 4D black holes
- even for compactification on a six-dimensional torus -  is
nontrivial. Moreover, the generalization of the existing proposals
for $\dim \CH(\gamma)$ to generic Calabi-Yau 3-folds will be much
more subtle and intricate.

For all these reasons, we are also willing to entertain the

\bigskip

\noindent {\bf WSV Conjecture.}
 ({\it Weak SV conjecture}): If $\{ \gamma_n \}$ is a
big sequence then \eqn\wkfrm{
 \lim_{n \rightarrow \infty}   ~~ {\log S_{\rm micro}(\gamma_n)
 \over \log z_n^2} = 1.
}

\bigskip

Of course, the SSV conjecture implies the WSV conjecture, but the
converse is false. The motivation behind the formulation of the
WSV conjecture is the following. It might be that for large
$\gamma$,
 the microscopic entropy actually
behaves like \eqn\distribution{ S_{\rm micro}(\gamma_n) \sim \pi
z_n^2 (\log z_n)^\alpha, } where $\alpha$ is a random variable,
 chosen using a
measure on $H_3(X;\IZ)$. Presumably the distribution would become
more sharply concentrated around $\alpha=0$ as $\gamma$ becomes
larger.
 As far as we know there
is nothing wrong with this idea, and from some viewpoints it even
seems likely. For example,
 the existence of BPS states can depend
on arithmetic properties of $\gamma$. So there might well be  a
large amount of ``scatter'' and ``noise'' in the behavior of $\dim
\CH(\gamma)$ as $\gamma \rightarrow \infty$. (See Fig. 1 in
section 5 below for an illustration of the kind of scatter we
mean.) In this case the Bekenstein-Hawking formula  would not be
the leading term in a systematic expansion, but would only hold in
some average sense.

If  the dimensions $\dim \CH(\gamma)$ indeed behave in the way
just suggested, then it is quite possible that the limit \wkfrm\
does not exist in the standard sense. In such a situation it is
more appropriate to use the lim-sup, (also denoted as
$\overline{\lim}$)
 which {\it always exists}
for any sequence of real numbers. \foot{We recall the definition.
If    $a_n$ is any sequence of real numbers let $b_n = \{ a_n,
a_{n+1}, \dots \}$. Then let $c_n:= {\rm sup} ~ b_n $ be  the
smallest constant   bounding $b_n$ from above. Plainly, the $c_n$
form
 a strictly decreasing sequence of real numbers.
We define  $\overline{\lim} ~a_n = \lim ~ c_n $. } Thus we could
replace the limits in the WSV conjecture by $\underline{\lim} $ or
$\overline{\lim}$. We refer to these as the $\underline{WSV}$ and
$\overline{WSV}$ conjectures, respectively.

There are yet other inequivalent formulations of the SV
conjecture.

%
%

\newsec{The height conjecture}

\subsec{Sharpening the height conjecture}

The main conjectures of \arthatt\  (conjectures 8.2.1, 8.2.2, and
8.2.3) posit that the attractor variety  $X_\gamma$  can be
defined as an arithmetic variety
 over some  number field $K_\gamma$.
This was verified in \arthatt\ in some special cases, for example
when $X$ is a free quotient of $K3 \times E$ (where $E$ is an
elliptic curve), or when $X$ is a complex torus. In a much more
speculative section,  \arthatt\ also suggested a possible
connection between the Faltings height for a metrized line bundle
${\rm ht}(X_\gamma;K_\gamma) $ and the entropy. We want to
investigate the consequences of these latter conjectures.

We begin by making the ``height conjecture'' more precise. The
first point to note is that if an attractor variety $X_\gamma$
indeed satisfies the attractor conjectures, then it might
nevertheless admit several different arithmetic ``models'' with
quite different arithmetic properties. First of all, changing the
field of definition of $X$ can alter the arithmetic properties. We
will not investigate the issues of ``base change'' systematically,
although we  note that the Faltings height does stabilize under
field extension \silvermanag. A second  ambiguity,  of more direct
relevance to what follows, is that two varieties $X,X'$ over a
field $K$ might be isomorphic as varieties over an algebraic
closure $\bar K$ (such as $\bar K = \IC$ for a number field), but
fail to be isomorphic as varieties over $K$. We will quote some
relevant examples in section 3.3 below.

As with the SV conjecture, there are various inequivalent precise
formulations of the height conjecture:

\bigskip
\noindent {\bf Strong Height (SH) Conjecture}: Assume a family of
arithmetic attractor varieties $(X_{\gamma_n},K_{\gamma_n})$
corresponds to a big sequence, that is,
   $z_n:=\vert \CZ(\gamma_n) \vert \rightarrow
\infty$. Then there exists a  finite positive constant $\kappa$
(possibly depending on the family) and a choice of model for
$X_{\gamma_n}$ such that if  ${\rm ht}(X_\gamma;K_\gamma) $ is the
Faltings height for the metrized line bundle provided by the
Calabi-Yau data, then \eqn\htconj{ \exp \bigl[ {1 \over  \kappa}
{\rm ht}(X_{\gamma_n};K_{\gamma_n})] = S_{\rm micro}(\gamma_n) +
o(\vert z_n \vert^2 ) . }

Alternatively, with the same hypotheses, we can formulate the

\bigskip
\noindent {\bf Weak Height (WH) Conjecture}: \eqn\htconj{
 \lim_{n \rightarrow \infty }  ~~
{ {\rm ht}(X_{\gamma_n} ;K_{\gamma_n}) \over \log[  S_{\rm
micro}(\gamma_n)]  }= \kappa. }

Of course, since not much is known about either numerator or
denominator in \htconj, it is prudent to allow alternative
formulations $\overline{WH}$, and $\underline{WH}$ of the
conjectures, with constants $\overline{\kappa}$ and
$\underline{\kappa}$, respectively. If the WH or SH conjectures
turn out to be true, then the  extent to which the constant
$\kappa$ depends on the family of charges will become an
interesting question, as will become apparent in section five.

Let us return to the ambiguities of base-change and choice of
model for $X_\gamma$. Regarding base change, our  hope is that the
choice of base field, while necessary to define the heights,
should do no more than change the value of the constant $\kappa$
in the height conjectures.

Regarding the choice of model, the attractor equation of
supergravity only specifies the attractor variety $X_\gamma$ as a
variety over $\IC$. If $X_\gamma$ can be defined over a number
field $K_\gamma$, then since  there can be inequivalent models
over $K_\gamma$ we must ask if the choice of model has any
relevance for the physics of string compactification on $X$. A
physicist's natural reaction to this question would be that the
choice of model should be irrelevant, by general covariance.
\foot{Roughly speaking, general covariance refers to invariance of
a theory of gravity under $\CC^\infty$ diffeomorphisms. A blind
application of this principle would also suggest that complex
structure is physically irrelevant (which is hardly the case). The
main thesis of this exploratory paper is that not only complex
structure, but even arithmetic structure is physically
significant.} However, as seen in \arthatt, the arithmetic of the
number fields associated to attractor varieties {\it is} related
to  such physical quantities as the BPS mass spectrum. Whether or
not the choice of arithmetic model is also of physical relevance
remains to be seen. In fact, the height conjectures above are the
first instance, of which we are aware, in which such a choice
really matters. This raises the interesting question of whether
physics indeed selects a distinguished arithmetic model. \foot{A
good place to start thinking about this might be Witten's linear
sigma model formulation of CY sigma models, where a definite
choice of projective model (albeit over $\IC$)  is made by the
quantum field theory \wittenls.}

\subsec{A special class of attractor varieties}

Some weak evidence for the height conjecture was given in
\arthatt\ in the case when $X$ is a complex torus. In this case it
was shown that $X_\gamma$ is   isogenous to a product of 3
elliptic curves with complex multiplication: \eqn\isogeny{
X_\gamma \sim  E_{\tau(\gamma)} \times E_{\tau(\gamma)}  \times
E_{\tau(\gamma)} } where $E_\tau := \IC/(\IZ + \tau \IZ)$ and
$\tau(\gamma) = i \sqrt{I_4(\gamma)}$.
 Here $I_4(\gamma)$ is an integral quartic polynomial
on $H_3(X;\IZ)$ related to the quartic  $E_{7,7}$ invariant (see
\arthatt\ for precise definitions). Moreover, one easily verifies
in this example that: \eqn\toreye{ \vert \CZ(\gamma) \vert^2 =
\sqrt{I_4(\gamma)} , } while the supergravity analysis
 shows that indeed
$S_{\rm sugra}(\gamma) = \pi  \sqrt{I_4(\gamma)}$
\kalkol\enneightbh\fm.

Equation \isogeny\ is only a statement about isogeny classes. In
order to estimate the height it is more convenient (but not
absolutely necessary) to pin down the attractor variety exactly.
We  will eliminate the unkown isogeny by choosing ``diagonal
charges.'' That is, using  equations $(6.2)-(6.5)$ of \arthatt,
with  $P^{ij} = p^i \delta^{ij}, Q_{ij} = q_i \delta_{ij}$ in the
notation of that paper, we may choose lattice vectors  $\gamma\in
H^3(X;\IZ)$ depending on eight integers $r, p^1,p^2,p^3, s,
q_1,q_2$, and
 $q_3$ such that we have the equality
\eqn\isogenyp{ X_\gamma =  E_{\tau_1} \times E_{\tau_2}  \times
E_{\tau_3}, } where  now we have \eqn\speceyef{ I_4(\gamma)   = 2
\bigl[   ( \sum_{i=1}^3 p^i q_i )^2 -
 \sum_{i=1}^3 (p^i q_i )^2 \bigr] - (r s + \sum_{i=1}^3 p^i q_i )^2 + 4 \bigl(
r q_1 q_2 q_3 - s p^1 p^2 p^3 \bigr) } and, defining $D= -I_4 <0$,
$\tau_i$ are given by: \eqn\heegner{ \tau_1 = { 2 \bigl( p^1 q_1
- p^2 q_2 - p^3 q_3 - r s\bigr) + \sqrt{D} \over 4( p^2 p^3 + r
q_1) } } with cyclic permutations on $123$ giving the formulae for
$\tau_2, \tau_3$.

Now let us discuss some arithmetic associated with the elliptic
curves $E_{\tau_i}$. In general, if $\tau$ is of the form $(-b +
\sqrt{D})/2a$ for integral $a, b, D$, with $D<0$,  then it follows
that $\IZ + \tau \IZ$ is a proper fractional ideal for some  order
$\CO$ of the field $K_{D}:= \IQ(  \sqrt{D})$. To
 be more precise,
let the minimal polynomial of $\tau$ be $A \tau^2 + B \tau + C =0$
where $A,B,C$ are integral. (Note: it need not be true that $a=A$,
or $b=B$.) Then by \cox, Lemma 7.5,  $\IZ + \tau \IZ$ is a proper
fractional ideal for the order $\CO = \IZ + A \tau \IZ$ of
$K_{D}$. In general, the conductor of the order $\CO$ will be
larger than one. By \cox, Theorem 11.1, the value of the modular
function $j(\tau)$ generates
 the ring class field
of this order, i.e., the ring class field is
 $K_{D}(j(\tau))$.  It follows from \isogenyp\
that $X$ is at least defined over the compositum of three such
ring class fields (it might well be defined over a smaller field).

We would like  to eliminate some of the complications of general
ring class fields and work instead with the Hilbert class field
$\widehat{K}_D$ of $K_D$. This will simplify the height
computation below. As explained at length in \cox, the distinct
 ideal classes of the ring of integers
$\CO(K_D)$ of $K_D$ have representatives $\IZ + \tau_k \IZ$. Here
$k=1,\dots, h(D)$, and $h(D)$ is the class number of $K_D$. The
complex numbers $\tau_k$ have the form $\tau_k = (-b_k +
\sqrt{D})/2 a_k$ and are  the solutions in the upper half-plane to
$a_k \tau^2 + b_k \tau + c_k=0$ where $a_k x^2 + b_k xy + c_k y^2$
runs over representatives of  inequivalent primitive binary
quadratic forms of  discriminant $D$. We can choose
representatives in the standard ``keyhole'' fundamental domain for
 $PSL(2,\IZ)$.
We will refer to the $\tau_k$ as ``Heegner points.'' One of the
beautiful statements of the theory of complex multiplication is
that for any $k$, $j(\tau_k)$ is an algebraic integer and
$\widehat{K}_D \cong K_D(j(\tau_k))$.

Now let us consider some explicit charge vectors $\gamma$. First
we take  $p= p^1 = p^2 = p^3$, and $q= q_1 = q_2 = q_3$ so that we
only have to work with a single elliptic curve with modular
parameter
  $\tau= \tau_1 = \tau_2 = \tau_3$ given by
\eqn\simpltau{ \tau = {-2 (pq + rs) +   \sqrt{D} \over  4 (p^2 + r
q) }. } Here $D$ reduces to $-I_4$, with $I_4$ given by
\eqn\specialdee{ I_4(p,q,r,s) = 12 p^2 q^2 + 4 (r q^3 - s p^3) -
(3 pq + rs)^2 . } Next we choose $p,q,r,s$ so that $\tau$ is one
of the Heegner points. We will content ourselves with finding
charge vectors $\gamma_D$ which yield the principal class (the
trivial class). This is given by \eqn\halfheeg{ \tau = {-1 +
\sqrt{D} \over 2} } if $D=1 ~ \mod ~4$ and $\tau = \sqrt{D}$ if
$D=0 ~ \mod ~ 4$. Specifically, if we
 choose $r=0$ and $s=3+4 D$, then with $p=q=4$, we get
\halfheeg. If $p=1,q=2,$ we get \eqn\heeg{\tau=-1+\sqrt{D}, }
which is clearly $PSL(2,\IZ)$-equivalent to $\tau = \sqrt{D}$.

Presumably, other charges $\gamma$ can lead to other ideal classes
in $K_D$, but we will focus on the above sequence of charges in
this paper, and will call them $\gamma_D$. As explained in the
next section, we will take $\widehat{K}_D$ as the field of
definition of the special attractor varieties associated with the
charges $\gamma_D$.

\subsec{Silverman's formula for the height of an arithmetic
elliptic curve}

Now that we have focused on the attractors $X_{\gamma_D}$ let us
compute their height. Since $X_{\gamma_D}$ is a product of 3
elliptic curves we have \eqn\relheit{ {\rm ht}(X_{\gamma_D} ;
\widehat K_{D} ) = 3 {\rm ht}( E_{\tau(\gamma_D)} ;
\widehat{K}_{D})  , } and therefore we need only compute the
height of an  elliptic curve. This has been done in a very
explicit way by Silverman in chapter 10 of \silvermanag. We now
review
 his formula. In the next section we
apply the formula to our considerations.

To begin, we must review a few standard definitions. (See any
textbook on elliptic curves, for examples
\cassels\husemoller\silveraec.) Let $K$ be a  field.
%
%
An elliptic curve $E/K$ can be given by a Weierstrass model
\eqn\genmodel{ y^2 + a_1 xy + a_3 y = x^3 + a_2 x^2 + a_4 x + a_6
} where $a_i \in K$. Two equivalent Weierstrass models for the
same curve over $K$ are related by a change of coordinates
\eqn\chgvars{ \eqalign{ x' & = u^2 x + r \cr y' & = u^3 y + s u^2
x + t \cr} } with $u \in K^*, r,s,t\in K$.

We introduce the discriminant $\Delta$
 through the standard definitions:
\eqn\beecee{ \eqalign{ b_2 & := a_1^2 + 4 a_2\cr b_4 &:= a_1 a_3 +
2 a_4\cr b_6 & :=a_3^2 + 4 a_6\cr b_8 & := a_1^2 a_6 - a_1 a_3 a_4
+ 4 a_2 a_6 + a_2 a_3^2 - a_4^2\cr c_4 & := b_2^2 - 24 b_4 \cr c_6
& := -b_2^3 + 36 b_2 b_4 -216 b_6\cr \Delta & := -b_2^2 b_8 - 8
b_4^3 - 27 b_6^2 + 9 b_2 b_4 b_6\cr} } If the elliptic curve is
nonsingular then $\Delta\not=0$ and we can define the
$j$-invariant: \eqn\jayinv{\ j  := c_4^3/\Delta. } When the
characteristic is not $2$ or $3$ it is useful to note that $2^6
3^3  \Delta = c_4^3 - c_6^2$.

Under the change of variables \chgvars\ we have: \eqn\chgvari{
\eqalign{ c_4' & = u^4 c_4 \cr c_6' & = u^6 c_6 \cr \Delta' & =
u^{12} \Delta\cr j' & = j \cr} } Curves with the same
$j$-invariant are isomorphic over the algebraic closure of $K$,
but need not be isomorphic over $K$.

Now let $K$ be a number field and $E$ an elliptic curve over $K$.
The field $K$ has a set of valuations. These consist of
archimedean valuations (``the places at infinity'') and
nonarchimedean valuations (``the finite places''). The places at
infinity correspond to inequivalent embeddings of $K$ into $\IR$
or $\IC$. In our example below all the embeddings will be complex
embeddings $\psi_i:K \hookrightarrow \IC$, so we henceforth assume
this is the case for $K$.
 Under
these embeddings $E(\IC)$ will be isomorphic to $\IC/(\IZ + \tau_i
\IZ)$. The finite places correspond to valuations labelled by
 the different prime ideals in $K$.
Let ${\bf p}$ be a prime ideal in $K$.
\def\bp{{\bf p}}
We can then consider the curve $E_\bp$
 over the $\bp$-adic completion $K_\bp$
which is just given by the same equation for $E$ as before,
 but now considered over this larger
field. Let $(\Delta)_\bp$ be the discriminant of $E_\bp$ (this is
an ideal in $\CO(K_{\bp})$). A {\it minimal model for $E_\bp$} (at
$\bp$)
 is obtained by
making changes of coordinates of the form \chgvars\
 (with
the field now taken to be $K_\bp$) such that the power of $\bp$
dividing $(\Delta)_{\bp}$ is the smallest possible. For the
minimal model let us say ${\bf p}^{n_{\bp}}\vert (\Delta)_{\bp}$,
but $ {\bf p}^{n_{\bp}+1 }\not\vert (\Delta)_{\bp}$.
  In particular, if
$E$ is smooth at $\bp$ then $n_{\bp}=0$,
 and $n_{\bp}>0$
only at those primes which divide $(\Delta)$ in the ``global
model'' of  $E$. Taking the product over all the finite places of
$K$ defines the {\it minimal discriminant} of $E$ as an ideal in
$K$: \eqn\defofmind{ \CD_{E/K} :=\prod_{\bp} \bp^{n_\bp}. }

We can now state Silverman's formula for the height of $E/K$
(Proposition  1.1, p. 254 in \silvermanag.) \foot{We will not
define the height. See, for examples, \faltings\silvermanag\lang\
for definitions. For expository discussions see \bloch\mazur.
Curiously, the Faltings height has appeared before in string
theory. See, for example,  \bost, and related papers. We do not
understand any connection to this  work.} The formula  for the
height involves  the
 sum of the
contributions from the finite and infinite places:
\eqn\silverform{ 12 [K:\IQ] {\rm ht}(E;K) = \log \vert
N_{K/\IQ}(\CD_{E/K}) \vert - 2\sum_i \log\biggl[ (\Im \tau_i
)^6\vert \eta^{24}(\tau_i)\vert (2\pi)^{-12}  \biggr] } where
$N_{K/\IQ}$ is the norm of the ideal,
 the second term is the sum over inequivalent
complex embeddings of $K$, and $\eta(\tau)$ is the Dedekind
function.

Now, to apply \silverform\ we must {\it choose}
 a field
of definition and a Weierstrass model for our attractor varieties
with $\tau$ given by \halfheeg\ or \heeg. This will introduce some
arbitrariness into our discussion. We motivate our choice as
follows. If $j\not= 0, (12)^3$ then one can always write a model
for an elliptic curve over $\IQ(j)$ with invariant $j$ by taking:
\eqn\curvejay{ y^2 =4 x^3  - {27 j \over j - 1728} (x+1). } Thus,
one obvious model for the attractor varieties is obtained by
making a transformation of the form \chgvars\ to get \eqn\overarr{
\eqalign{ y^2 & = 4 x^3 - g_2 x - g_3 \cr g_2& = 27 j (j- (12)^3)
\cr g_3 & = 27 j (j-(12)^3)^2\cr \Delta & = g_2^3 - 27 g_3^2 = 2^6
\cdot 3^{12}\cdot j^2 (j-(12)^3)^3. \cr} } Note that we are taking
the coefficients
 in the ring of
integers  $\CO_{\widehat{K}_D}$. Note too that the Hilbert class
field is not the minimal field of definition of this curve. We
could take $\IQ(j(\tau))$. However, this field has many different
conjugates inside the Hilbert class field, depending on which
Heegner point $\tau$ is taken.  Thus we find the Hilbert class
field more natural.

It is important to note that \overarr\ is not the only Weierstrass
model we could choose. We can illustrate the kinds of ambiguities
we face in the following two simple sets of examples. Consider two
families  of elliptic curves over $\IQ$ labelled by an integer $n$
\eqn\expleone{ \eqalign{ y^2 & = x^3 + n\cr y^2 & = x^3 + nx. \cr}
} The first family has $j=0$, $c_4=0, c_6 = -2^5 3^3 n, \Delta = -
2^4 3^3 n^2$ and complex multiplication by $\IZ[{1+\sqrt{-3}\over
2}]$; the second family has $j=2^6 3^3, c_4=-2^3 3 n, c_6=0,
\Delta = -2^6 n^3$ and complex multiplication by $\IZ[\sqrt{-1}]$.
Nevertheless, if we consider two curves for $n_1,n_2$ then, if
$n_1/n_2$ is not a sixth power in the first  example or a  fourth
power in the second example,   the curves are inequivalent over
$\IQ$ and only become equivalent over extensions of degree 6 and
4, respectively. Moreover, {\it the minimal discriminants depend
on the choice of $n$. } Less trivial examples of inequivalent
Weierstrass models (with $j\not=0,1728$) can be gleaned from
\cremona.

%
%

\subsec{Bounding the height}

We now specialize Silverman's formula further and put a bound on
the height so that we can test the height conjecture.

In order to specialize Silverman's formula, we note that the
values of the $j$-function at the different Heegner points
$\tau_k$ define the distinct embeddings of
$\widehat{K}_D\hookrightarrow \IC$, so we may  rewrite Silverman's
formula for the height of \overarr\ as \eqn\heightcrve{ \eqalign{
24 h(D) h(E_\gamma /\widehat{K}_D ) & = \log \vert
N_{\widehat{K}_D/\IQ}(\CD_{E_\gamma/\widehat{K}_D })\vert +
\sum_{k=1}^{h(D)} \CR(\tau_k)\cr \CR(\tau) & := -2\ \log\biggl[
(\Im \tau )^6\vert \eta^{24}(\tau)\vert (2\pi)^{-12}  \biggr].\cr}
} The   sum in \heightcrve\ is over representatives for the ideal
class group. In general we will write: \eqn\defaverg{ \langle
f(\tau)\rangle_D := {1 \over h(D)} \sum_{k=1}^{h(D)} f(\tau_k) }
for any function $f$. For $f(\tau) = \CR(\tau)$ we denote $R(D) :=
\langle \CR(\tau)\rangle_D$. Thus, for our special charges
$\gamma_D$
 \eqn\htarr{ {\rm ht}(X_{\gamma_D} ;\widehat K_{D} ) =
{1 \over  8} \biggl[ R(D) + {\log \vert
 N_{\widehat{K}_D/\IQ}(\CD_{E/\widehat{K}_D })\vert \over  h(D) }
  \biggr]
.} 
 The term  $R(D)$ in \htarr\
 only depends on $\gamma$
through the $E_{7,7}$ invariant
 $D = - I_4(\gamma)$.
As we have discussed above, the minimal discriminant can depend on
the choice of Heegner point defining the curve \overarr, and even
on the choice of $\widehat K_D$-isomorphism class
 for the curve.  We have, somewhat
arbitrarily, chosen \halfheeg\heeg\ corresponding to the principal
class and moreover have chosen the Weierstrass model \overarr.

Now we can put some useful bounds on the height. The minimal
discriminant is a subtle object and is hard to estimate, even for
a CM curve. For some CM curves its  norm can be as small as $1$
(over $F=\IQ(j)$ under certain conditions on $D$ -- see
\rohrlich). In \bgross  it is calculated for an interesting family
of elliptic curves. It is quite possible that further results for
the model \overarr\ can be obtained from the deep work of Gross
and Zagier \gzsm. Nevertheless, while it is subtle,  we {\it do}
know that in the model \overarr\ it is
 a certain integral ideal in $\widehat{K}_D$
which divides the principal ideal $(\Delta)$ explicitly calculated
in  \overarr. {}From this we may derive some easy inequalities, as
in \arthatt: \eqn\inequals{ {1 \over  8}   R(D)
  \leq {\rm ht}(X_{\gamma_D}, \widehat K_D)
\leq {6 \over  8}   R(D)  + \langle f(\tau) \rangle_D } where
\eqn\giveff{ f(\tau) = 6 \log\bigl[ (\Im \tau)^4 \vert E_4(\tau)
\vert^2 \bigr] + 6 \log\bigl[ (\Im \tau)^6 \vert E_6(\tau) \vert^2
\bigr] + const. } The first inequality in \inequals\ does not
depend on the choice of Weierstrass model, but the second does..

The function $f(\tau)$ grows like $60 \log(\Im \tau)$ for $\tau
\rightarrow i \infty$ and is therefore square integrable in the
Poincar\'e measure. It follows from a
 theorem of W. Duke
\duke\ that the average $\langle f(\tau) \rangle_D$ converges to
\eqn\average{ \langle f(\tau) \rangle_D \rightarrow {3 \over  \pi}
\int_{\CF} {dx dy \over  y^2} f(x+iy) } as $D \rightarrow -
\infty$. In particular, the integral and hence the limit is
finite.

\newsec{Summary of some analytic number theory}

In sections 4.1 and 4.2
 we summarize some  well-known facts and
definitions from analytic number theory. In section 4.3 we
summarize some more technical facts needed in section 5.

\subsec{$L$-functions}

\defn{4.1: Legendre-Jacobi-Kronecker symbol.}
 This is the unique, real, nontrivial
Dirichlet character of modulus $D$. Its value for $n$ is denoted
$\bigl( {D \over  n} \bigr)$.

The LJK symbol $\bigl( {D \over  n} \bigr)$ can be computed for
$D, n\not= 0$ as follows. First of all, it is completely
multiplicative in both arguments: \eqn\multiplv{ \bigl( {D_1 D_2
\over  n} \bigr) = \bigl( {D_1 \over  n} \bigr) \bigl( {  D_2
\over  n} \bigr) \qquad\qquad \bigl( {D\over  n_1 n_2 } \bigr) =
\bigl( {D \over  n_1} \bigr) \bigl( {  D\over  n_2} \bigr) } Thus
it suffices to give its value for $n=-1$ and for $n$ prime:

1. If $n=-1$ then \eqn\munus{ \eqalign{ \bigl( {D \over  -1}
\bigr) & =  +1 \qquad D\geq 0 \cr & =  -1 \qquad D < 0 . \cr}}

2. If $n=2$ then \eqn\kronk{ \eqalign{ \bigl( {D \over  2} \bigr)
& = 0 \qquad\qquad\qquad D ~~ {\rm even} \cr & = (-1)^{(D^2-1)/8}
\qquad D ~~ {\rm odd}. \cr} }

3. Finally, if $n=p$ is an odd prime then $\bigl( {D \over  p}
\bigr)$ is the Legendre symbol, i.e. \eqn\leng{ \eqalign{ \bigl(
{D \over  p} \bigr) & = 0 \quad\qquad\qquad if \qquad D=0~ \mod
~p\cr & = +1 \qquad\qquad if \qquad D=x^2 \mod ~p, ~ {\rm for}\
{\rm some} \ x\not=0 \cr & = -1 \qquad\qquad if \qquad D\not=x^2
\mod ~p, ~{\rm for} \  {\rm any} \  x \cr} }

One can check that if $D=0,1 \mod ~4$, (as in this paper), $\bigl(
{D \over  n} \bigr) $ only depends on the residue class $n ~ \mod
~\vert D \vert$.

\defn{4.2: Fundamental Discriminants} $D<0$ is called a fundamental
discriminant if it is the product of relatively prime factors of
the form $-4,8,-8$, or $(-1)^{(p-1)/2}p,p\ge 3$.

Equivalently, $D<0$ satisfies either (a.) $D=1~\mod ~4$ and  $D$
is squarefree, or  (b.) $D=0~\mod ~ 4$ ,  $D/4\not=1~\mod ~ 4$,
and $D/4$ is squarefree. These are the discriminants of quadratic
imaginary fields.

\defn{4.3: L-functions} $L$-functions of conductor
$D$ are by definition the infinite series \eqn\ellfun{ L(s,D) :=
\sum_{n=1}^\infty \bigl( {D \over  n} \bigr) n^{-s} . }

For example, choosing $D=1$ we get the Riemann $\zeta$ function.
Other examples of $L(s,D)$ are: \eqn\ellexpls{ \eqalign{ L(s,-3) &
=  \sum_{n=0}^\infty \left({1 \over (3n+1)^s} -
 {1 \over (3n+2)^s}\right) \cr
L(s,-4) & =  \sum_{n=0}^\infty \left({1 \over (4n+1)^s} -
 {1 \over (4n+3)^s}\right)   \cr
L(s,-7) & = \sum_{n=0}^\infty \biggl({1 \over (7n+1)^s} + {1 \over
(7n+2)^s} -{1 \over (7n+3)^s}+\cr
 & +  {1 \over (7n+4)^s}   - {1 \over
(7n+5)^s} - {1 \over (7n+6)^s} \biggr) \cr} }

The analytic properties of $L(s,D)$ are well known  \davenport.
The series is absolutely convergent for $Re(s)>1$ and admits an
analytic continuation as an entire  function of $s$ (for
$D\not=1$).  Moreover, for $D<0$ we may define \eqn\defxii{
\xi(s,D) := (q /\pi)^{(s+1)/2} \Gamma((s+1)/2) L(s,D). } Here $q$
is a positive integer defined to be  the minimal period in $n$ of
the function $\bigl( {D \over  n} \bigr) $.  In general it is a
positive integer dividing $D$, but in our case where $D$ is a
fundamental discriminant, we have  $q=|D|$.  That is, the
character $\bigl( {D \over  n} \bigr)$ is primitive with period
$|D|$ exactly when $D$ is a fundamental discriminant.

  It can be shown that
 $\xi(s,D)$ is an entire function  of complex order
one. Moreover the   zeroes of $\xi(s,D)$ come in complex conjugate
pairs. Thus $\xi(s,D)$ has a product formula \foot{This notation
is a little sloppy because one should include a convergence factor
for the infinite product. Alternatively, one can eliminate this by
grouping the factors with $\Im \rho \not=0$ into complex-conjugate
pairs (since the coefficients of the Dirichlet series are real.)}
\eqn\prodform{ \xi(s,D) = e^A \prod_{\rho} \bigl( 1 - { s \over
\rho} \bigr). }

\def\frac#1#2{{#1 \over #2}}
The completed function $\xi(s,D)$ has the functional equation
\eqn\functen{ \xi(s,D)=\xi(1-s,D), } which is valid only for
primitive characters ($q=\vert D\vert $).  Otherwise one must make
a modification to account for prime factors of $\vert D\vert/q$.
The  zeroes $\rho$, called  the {\it critical zeroes} of $L(s,D)$,
lie in the critical strip $0 \leq Re(s) \leq 1$.

It is known that $s=1$ is {\it not} a zero for any $D$, and in
fact \eqn\clssnum{ L(1,D) =   {2\pi h(D) \over    w(D) \vert D
\vert^{1/2} } . } (We will recall a proof of this below.) Here
$w(D)$ is the order of the group of units in $K_D$; for $D<-4$ we
have $w(D) = 2$.

\subsec{ The zeroes of $L(s,D)$}

The LJK symbol is completely multiplicative as a function of $n$.
Combining this with the prime factorization of integers one can
write {\it another} product  formula \eqn\eulerprod{ L(s,D) =
\prod_{\rm p~ prime } \biggl( 1- \bigl({D \over p}\bigr)
p^{-s}\biggr)^{-1} }
 for
the function $L(s,D)$. This is analogous to Euler's product
formula for $\zeta(s)$. While it is not immediately obvious, the
product formula \eulerprod\ for $L(s,D)$ encodes
 the structure of primes in
the field $K_D$.   Thus, comparing the two product formulae
\prodform\ and \eulerprod\  gives information about primes.
Accordingly, the nature of the zeroes of $L(s,D)$ are of interest
in understanding the arithmetic of $K_D$.

This astute remark leads to the central problem of analytic number
theory,
  the generalized Riemann hypothesis. It is
``generalized'' because we are considering the family of functions
$L(s,D)$; henceforth we use the abbreviation GRH. \foot{In this
paper GRH always stands for the hypothesis for Dirichlet
$L$-functions, not for elliptic curve $L$-functions.} A
formulation of the GRH, suitable for our present purposes, is the
conjecture that the critical zeros of $L(s,D)$ all lie on the line
$Re(s) = \half$. The GRH is of course very difficult,
 so various
sub-problems have been studied. One of these concerns the possible
existence of critical zeroes near $s=1$. These are known as
``Landau-Siegel zeroes'' (LSZ's). Their existence would falsify
the GRH. More precisely:

\defn{4.4: Landau-Siegel zeroes.} Choose a constant $c>0$.
A ``{\it Landau-Siegel zero for $c,D$}''
  is a real zero $s=\beta$ of $L(s,D)$ with
\eqn\lsz{ 1 - { c \over  \log \vert D \vert} < \beta < 1 . } See
\davenport\ (chapter 21) for equivalent formulations of the
Landau-Siegel zero.

It is possible to explicitly compute an effective constant $c_2>0$
such that the neighborhood of radius $c_2/\log\vert D\vert$ about
the point $s=1$ contains at {\it most} one zero of $L(s,D)$.  This
rules out complex zeroes in this neighborhood since they come in
conjugate pairs. However, the important challenge remains to
eliminate the real zeroes. That is,
 to find an {\it effective}
constant $c_1$ such that there are no Landau-Siegel zeroes for
$c_1$  and any $D$, whether positive or  negative.  In this paper
we only study $D<0$.

\subsec{The large $\vert D \vert$ behavior of $L(1,D)$ and the
zeroes of $L(s,D)$}

In this section we explain  how the zeroes of $L(s,D)$ are related
to  the
 large $\vert D \vert$ behavior
of \eqn\llp{ \lambda(D) := {L'(1,D) \over  L(1,D) } , } where the
derivative is with respect to $s$. We will begin with some
heuristic remarks, and conclude with some precise estimates
(Theorems 4.1 and 4.2). These estimates will be useful in the next
section.

The relation between the ``size'' or rate of growth of
$\lambda(D)$ as $\vert D \vert \rightarrow \infty$ and the
existence of zeroes near $s=1$ follows from the key identity
\eqn\logdiff{ \frac{\xi'}{\xi}(s,D)=\frac{1}{2}\log(|D|/\pi)+
\frac{1}{2}\psi((s+1)/2)+\frac{L'}{L}(s,D) =\sum_{\rho\mid
\xi(\rho,D)=0} \frac{1}{s-\rho}. } Here \eqn\digamma{
\psi(x):={\Gamma'(x) \over  \Gamma(x)} } is the digamma function.
Equation \logdiff\ follows immediately from logarithmic
differentiation in $s$ of  the product formula \prodform.

It follows from \logdiff\ that
 $\frac{L'}{L}(1,D)$ can  grow rapidly with
$\vert D \vert$
 if $L(s,D)$ has some zero $\rho$ which is
``close'' (as a function of $\vert D \vert$) to the point $s=1$.
On the other hand, zeroes further away from $s=1$ have less of an
impact on the  sum   in \logdiff; their effect is   governed by
their ``density of states.'' It is only the lower-lying zeroes
which are important for the size of $\lambda(D)$. Therefore, if we
can say $\lambda(D)$ is ``small'' then we will have checked a
prediction of the GRH, at least in some neighborhood of $s=1$. The
smaller the bound on $\lambda(D)$ the larger the zero-free region
around $s=1$.

We will now explain a trivial bound on the growth of $\lambda(D)$
obtained from bounding the numerator and denominator separately.
Our argument is sloppy and far from optimal, yet it does explain
in simple terms the overall reason why the Landau-Siegel zero is
related to a small value of $L(1,D)$.

We first bound the numerator by showing there is a constant $C$
such that \eqn\tbndi{ \vert L'(1,D)\vert \leq C (\log|D|)^2. } We
prove this as follows. By  \ellfun\ \eqn\sumi{
L'(1,D)=-\sum_{n=1}^{|D|} (\frac{D}{n})(\log n)n^{-1}
        - \sum_{n=|D|+1}^\infty (\frac{D}{n})(\log n)n^{-1}  .
} The first sum is trivially \eqn\sumii{ \CO(\sum_{n=1}^{|D|}
(\log n)n^{-1}) =\CO((\log|D|)^2). }
  The second sum
can easily be bounded by $\CO((\log|D|)^2)$ using partial
summation, since $\frac{\log n}{n}$ changes slowly and
\eqn\sumiii{ \sum_{n=k}^{k+D} (\frac{D}{n})=0, k=1,2,3,\ldots }
 (Roughly speaking,
this means there is a lot of cancellation due to the oscillation
of the Dirichlet character which modulates $(\log n)n^{-1}$.)

The bound $L'(1,D)= \CO((\log \vert D \vert)^2)$ can be improved.
Nevertheless, the intuition is that the fraction $\lambda(D)$ can
be extremely large only if its denominator is extremely small.
Therefore we now   bound the denominator. A trivial bound comes
from \clssnum.  Since $h(D) \geq 1$ it follows  that \eqn\tbndii{
{1 \over  L(1,D) } = {\sqrt{\vert D \vert } \over  \pi h(D)} \leq
{\sqrt{\vert D \vert } \over  \pi  }. } Of course, $h(D)$ {\it
does} grow with $D$ so this estimate is far from optimal!
\foot{Note that if $h(D)/\vert D\vert^{1/2}$ is large there cannot
be a zero near $s=1$, while if $h(D)/\vert D\vert^{1/2}$ is small
there might be a nearby zero. This is the essence of the
connection to the class number problem.
 Indeed, the
work of Goldfeld and Gross-Zagier provides better bounds on
$1/L(1,D)$, reverses the logic of the present discussion,  and
proves a growth rate of $h(D)$ for $\vert D\vert$ increasing. This
provides a solution to Gauss' problem of finding which imaginary
quadratic fields have a given fixed class number. }

Taken together, \tbndi\ and \tbndii\  already show that
\eqn\tbndiii{ \vert \lambda(D) \vert \leq C' (\log \vert D
\vert)^2 \sqrt{\vert D \vert } . } {}From \logdiff, the inequality
\tbndiii\ rules out zeroes at ({\it roughly}) a distance of order
$$(\log \vert D \vert)^{-2} \vert D \vert^{-1/2}$$ from $s=1$. Of
course, the sizes of $L'(1,D)$ and $L(1,D)$ are related to each
other (by integration of $L'(s,D)$ in $s$), but it is surprisingly
difficult to improve on $\vert D \vert^{1/2}$-factor in the
trivial bound \tbndiii.

Let us now see what the GRH has to say about the growth of
$\lambda(D)$. First of all, Littlewood proved that the GRH implies
$1/L(1,D)=\CO(\log\log|D|)$ as $D\rightarrow-\infty$ \littlewood.
Moreover, he showed that the GRH implies
\eqn\lambdaupper{|\lambda(D)|=\CO(\log\log|D|), } and hence, in
particular, that the GRH forces
\eqn\lamoverlog{\lim_{D\rightarrow-\infty}\frac{\lambda(D)}{\log|D|}=0.
} In fact, one may produce sequences of $D$'s such that
$\lambda(D)$ grows roughly as $\log\log|D|$. All of this   is
evident from the following

\thm{4.1}[Miller -- in the appendix to this paper, \miller]. Then
\eqn\stvthrm{ \eqalign{ \limsup_{D\rightarrow -\infty} {\lambda(D)
\over \log\log \vert D \vert } & \ge \f 12 \cr
\liminf_{D\rightarrow -\infty} {\lambda(D) \over \log\log \vert D
\vert } & \leq  -\f 12 \cr} } where $\lambda(D)= L'(1,D)/L(1,D)$.
  These inequalities
 hold, in fact, if $D\rightarrow -\infty$ along
fundamental discriminants.

Of course, the GRH might actually be false! Evidently, there is a
good deal of room between the trivial bound \tbndiii\ and the
consequence \lamoverlog\ of the GRH. The elimination of
Landau-Siegel zeroes is thought to constitute an important
beachhead on the way towards understanding  the GRH.  The
following theorem gives a criterion to rule out the existence of
Landau-Siegel zeroes:

\thm{4.2} Suppose $D_n$ is a sequence of fundamental discriminants
with $D_n\rightarrow-\infty$.  Suppose furthermore that the limit
\eqn\lamlin{ \delta :=\overline{\lim}_{D_n\rightarrow-\infty}
\frac{\lambda(D_n)}{\log|D_n|} } is finite. Then there exists a
positive constant $c$, \eqn\cstarb{ c = \biggl(\half + \sup
\frac{|\lambda(D_n)|}{\log|D_n|} \biggr)^{-1} } such that there
are no Landau-Siegel zeroes for $c$ among the $D_n$.

\bigskip
\noindent {\it Proof:} Note that the zeroes $\rho$ are either real
or come in complex conjugate pairs. Writing $\rho = \beta + i
\gamma$ for the real and imaginary parts we have: \eqn\trivbdi{
\sum_{\rho} {1 \over 1- \rho} = \sum_\rho {1- \beta \over
(1-\beta)^2 + \gamma^2} } and since the zeroes are in the critical
strip, this is a sum of nonnegative terms. Thus, if $\rho_* =
\beta_*$ is a real zero of $L(s,D)$ then \eqn\trivbdii{ \eqalign{
{1 \over 1- \beta_*} & < \sum { 1 \over 1- \rho} \cr & = \half
\log(|D| /\pi) + \half \psi(1) + \lambda(D) \cr}, } where in the
second line we have used \logdiff. Recall that $\psi(x)$ is the
digamma function \digamma.

Since \trivbdi\ is positive and the constant $-\half \log \pi +
\half \psi(1) \cong - 0.86$ is negative it follows from \trivbdii\
that \eqn\lowerbnd{ -\half < {\lambda(D)\over \log \vert D\vert} }
for any $D$. Therefore, finiteness of
 \lamlin\ implies that
\eqn\supbnd{ \sup \frac{|\lambda(D_n)|}{\log|D_n|} < \infty, }

Now we use \eqn\anthrbnd{ \lambda(D_n) \leq  \left( \sup
\frac{|\lambda(D_n)|}{\log|D_n|}  \right) \log|D_n|. }
 Then \trivbdii\ shows that if $\beta_{*,n}$
is a real zero of $L(s,D_n)$, then \eqn\trivbdiv{ {1 \over 1-
\beta_{*,n} } < \left(\half+ \sup \frac{|\lambda(D_n)|}{\log|D_n|}
\right) \log \vert D_n\vert . } Now comparing with the definition
\lsz\ we see there is never a Landau-Siegel zero for $(c,D_n)$ for
$c=  \left( \half + \sup \frac{|\lambda(D_n)|}{\log|D_n|}
\right)^{-1}$.

An alternative formulation of this proof (which has the advantage
of naming the constant $c$ at the sacrifice of ignoring a finite
number of cases) is to use the literal definition of the lim sup:
for each $\epsilon>0$ one can find an $N(\epsilon)$ such that
\eqn\trivbdiii{ \biggl\vert \delta-{\rm sup}_{m\geq n}
\frac{\lambda(D_m)}{\log|D_m|} \biggr\vert  < \epsilon }
 for all $n\ge N(\epsilon)$.
Then \eqn\trivbdsup{ {1 \over 1- \beta_{*,n} } < (\half + \delta +
\epsilon) \log \vert D_n\vert } for $n\geq N(\epsilon)$, which
leads to the conclusion that there is no Landau-Siegel zero for
$(c,D_n)$ for $c=  (\half + \delta + \epsilon)^{-1} $ and $n\geq
N(\epsilon)$. $\spadesuit$

{\bf Remarks.}

\item{1.} Put differently,
if there are $c$'s $\rightarrow 0$ such that each has some
Landau-Siegel zero, then those discrimiants
$D_n\rightarrow-\infty$ satisfy $$\lim
\frac{\lambda(D_n)}{\log|D_n|}=\infty.$$

\item{2.}  The issue of effectiveness enters in the following ways.  In
the first argument, one might know $\delta$ exactly but not know
what $\sup \frac{|\lambda(D_n)|}{\log|D_n|} $ is -- we may know
the limiting behavior of this sequence but not be able to bound
how long it takes for this sequence to exhibit it.   In the second
argument, we may not be able to  compute $N(\epsilon)$
effectively.

\newsec{Black hole entropy and critical zeroes}

In this section we will explore come consequences for analytic
number theory of the various conjectures    of sections 2 and 3.

The main technical observation is that the height conjectures
have implications for the behavior  of $L(s,D)$ thanks to an
equation
 sometimes called the
Chowla-Selberg formula \weil. This is a consequence of class field
theory and the Kronecker limit formula. We next recall the
derivation of this formula.

 The derivation begins  with the  the nonholomorphic
Eisenstein series \eqn\nonhes{ E(\tau,s) := \sum_{n,m}' {y^s \over
\vert m \tau + n \vert^{2s}} } where $y = \Im \tau$ and the
  sum is over all pairs of integers $(n,m)\not= (0,0)$.
The sum is absolutely convergent for $Re(s)>1$, and admits a
meromorphic continuation to the entire $s$-plane. The (first)
Kronecker limit formula is the statement (see, e.g.,
\langelliptic, p. 273) \eqn\klimiti{ E(\tau,s) = {\pi \over  s-1}
-  2\pi \log \biggl[ 2 e^{-\gamma} y^{1/2} \vert \eta(\tau)
\vert^2 \biggr]  + \CO(s-1) } as $s \rightarrow 1$. Now, let
$\tau_k$ be a set of representatives of the $h(D)$ Heegner points
as in section 3. A simple application of class field theory shows
that \eqn\clssft{ w(D) \bigl({ \vert D \vert \over  4}
\bigr)^{s/2} \zeta(s) L(s,D) = \sum_{i=1}^{h(D) } E(\tau_i, s). }
We now take $s \rightarrow 1$ and compare the two sides of the
equation. Equating the residue of the pole leads to the standard
result \clssnum. Comparing the  constant terms leads to the
Chowla-Selberg formula \eqn\chowla{ R(D) = 12\biggl[  \log
\sqrt{-D} +   \lambda(D) +
  \log\bigl[ 8 \pi^2 e^{-  \gamma  } \bigr] \biggr]
} where $R(D)$ is defined by \heightcrve\defaverg.

\ifig\scatter{A plot of $\lambda(D)$ against $\vert D\vert $, as
$D$ runs through the negative prime fundamental discriminants with
$\vert D\vert < 2000$. Unfortunately $\log\log\vert D\vert$ grows
too slowly to see numerically the envelope predicted by Theorem
4.1. } {\epsfxsize3.0in\epsfbox{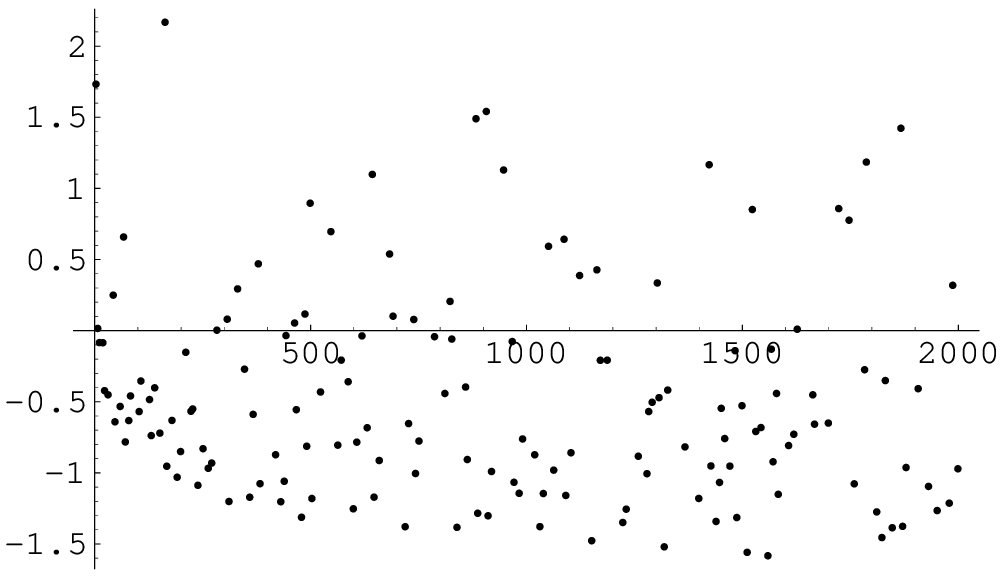}}

We are now ready to combine the conjectures of sections two and
three with the results from section four. The first point to make
is that the SSV and  SH conjectures
 are  ``probably'' {\it incompatible}. Let $D_n
\rightarrow -\infty$ be a sequence of fundamental discriminants.
Applying the
 SSV and
SH conjectures to the family of attractors $X_{\gamma_{D_n}}$
constructed in sections 3.2 and 3.3 we find \eqn\ssvsh{ {\rm
ht}(X_{\gamma_{D_n}},\widehat K_{D_n}) = \kappa \log [\pi
\sqrt{-D_n}] + \CO({\log [\sqrt{-D_n}]  \over\vert D_n
\vert^{1/2}},
 {1 \over
\vert D_n \vert^{\delta/4} } ). } This implies some very
interesting cancellation in \htarr\ for the following reason.
Roughly speaking, Theorem 4.1 says that  there exist real numbers
$\alpha_\pm$ with $\alpha_- < 0 < \alpha_+$, such that
$\lambda(D)$ can actually grow like $ \alpha_\pm \log(\log \vert D
\vert) $ for some sequence of $D$'s \miller. By \chowla\ this
means that $R(D)$ has scatter as illustrated in \scatter. For
large $\vert D\vert$ the envelope is of width
 $(\alpha_+ - \alpha_-)\log(\log|D|)$.
Suppose, for the moment, that the minimal discriminant term in
\htarr\ were absent. Then using the SH conjecture we would
conclude that $\kappa = 3/2$ and moreover, by choosing suitable
sequences of discriminants $D_n$, the charges $\gamma_{D_n}$ would
produce black holes with entropies  $S_{\rm micro}(\gamma_{D_n})$
that actually grow with $\vert D_n \vert$ like \eqn\biggrow{
S_{\rm micro}(\gamma_{D_n})\sim \pi \sqrt{\vert D_n\vert} (\log
\vert D_n \vert)^C } for various constants $C$ in the range
$\alpha_-  \leq C \leq \alpha_+ $. Moreover, Theorem 4.1 asserts
that there exist sequences  $\{D_n\}$ which actually realize the
extreme cases $\alpha_\pm$.

Now let us restore the  minimal discriminant term
\eqn\mindisct{{\log \vert
 N_{\widehat{K}_D/\IQ}(\CD_{E_\gamma/\widehat{K}_D })\vert \over  h(D) }
} in \htarr. As we have discussed, this depends on the choice of
Heegner point and the Weierstrass model.  It would be fascinating
(though we think unlikely) if for general sequences of charges
$\gamma_{D_n}$ with $D_n\rightarrow -\infty$ one could
systematically choose Weierstrass models such that the minimal
discriminant in \htarr\ fluctuates to match the changes in
$\lambda(D_n)$ in the way demanded by the SSV conjecture. Since we
think it is highly unlikely,  we conclude that SSV, and SH are
probably incompatible, as stated above. We could be more precise
if we had better information on the size of the minimal
discriminant $\CD_{E_\gamma/\widehat{K}_D }$.

Put differently, if one could choose families of attractor points
with a bounded value of  \mindisct\ then the strong height
conjecture implies that the distribution of \eqn\dimenrat{ {\log
\dim \CH(\gamma) \over  \pi \sqrt{I_4(\gamma)} } } plotted against
$\sqrt{I_4(\gamma)}$ would have a lot of scatter, similar to the
scatter in \scatter,  around the average value $1$.

Let us now assume WSV and $\overline{WH}$ (or WH). Then, from
\wkfrm\htconj\ and the first inequality in \inequals\ we get
\eqn\lowbndi{ \overline{\lim_n} ~~ {R(D_n) \over  \log \vert D_n
\vert^{1/2} } \leq 8\bar\kappa. } Then, from the Chowla-Selberg
formula we get \eqn\lowbndii{ \overline{\lim_n} ~~ {\lambda(D_n)
\over  \log \vert D_n \vert^{1/2} } \leq {8\bar\kappa -12 \over
12}  . } Now, using Theorem 4.2
  we arrive at the main statement in this paper:
The WSV conjecture and WH conjecture together rule out
Landau-Siegel zeroes for $D<0$. More precisely, we have

\bigskip
\thm{5.1} The WSV conjecture and the $\overline{WH}$ conjecture
together imply the following restrictions on Landau-Siegel zeroes:
For any sequence $\{ D_n \}$ of fundamental discriminants with
$D_n\rightarrow -\infty$, there exists a positive constant $c$
such that there are no Landau-Seigel zeroes for $(c,D_n)$.

\bigskip
We would like to remark that a similar result (with $\log \vert D
\vert$ replaced by $\vert D \vert^\epsilon$, for a positive
constant $\epsilon$) was proven unconditionally by Tatuzawa in
\tatuzawa. Tatuzawa's theorem, while weaker,  is sufficient for
many applications. Two novelties of the present discussion are
that we have $\log \vert D \vert$ rather than a power, and,
moreover, the constant $c$ is in principle computable from physics
(granted the
  WSV  and  $\overline{WH}$ conjectures).

Let us now turn the logic around and
 suppose that the correct configuration
of conjectures is WSV, WH, but that
 there are Landau-Siegel zeroes for $D<0$
(thus falsifying the GRH).  Then there would be sequences of
charges $\gamma_n$ with D-brane configurations of anomalously
large entropy compared to the Bekenstein-Hawking entropy.

Even more ambitiously, we can turn things around and assume the
``safest'' set of conjectures: GRH, WSV, and WH, and see what they
predict. {}From the GRH  we have: \eqn\rdeelimp{ \lim_{n
\rightarrow \infty} ~~ {R(D_n)  \over  \log \vert D_n \vert^{1/2}
} = 12 } Now, the  WSV and the $\overline{WH}$  conjecture imply
information on the mysterious minimal discriminants of the
elliptic curves \overarr. From \inequals\ we get \eqn\rdeelim{
\overline{\lim_n} ~~ {\log \vert N_{\widehat{K}_{D_n}/\IQ}
(\CD_{E_{\gamma_n}/\widehat{K}_{D_n} })\vert
 \over  h(D_n) \log ( \vert D_n \vert^{1/2}) }
= 8 \bar \kappa -12, } and in particular, $\bar \kappa \geq 3/2$.
Note that the second inequality in \inequals\  shows that $\bar
\kappa \leq 9$. The second inequality depends  on the choice of
Weierstrass model \overarr, so we are assuming our choice of
models is suitable for the conjecture. Of course, replacing
$\overline{WH}$ by $WH$ we replace $\bar \kappa$ by $\kappa$ and
$\overline{\lim}$ with $\lim$ and get a stronger ``prediction.''

Curiously, some examples support \rdeelim. As we mentioned before,
the minimal discriminant was computed in \bgross\ for a certain
family of curves  over $\widehat{K}_{D=-p}$ for $p=3~\mod ~4$, and
prime. Denoting these curves by $A(p)$, Theorem 12.2.1 of \bgross\
shows that $\CD_{A(p)/\widehat{K}_{D=-p}}= (-p^3)$, so the norm
over $\IQ$ is $(-p^3)^{2h}$ and we thus get $\bar \kappa = 3$ for
this family. On the other hand, there are also families of CM
curves with $N_{\widehat{K}_{D_n}/\IQ}(\CD) = 1$ \rohrlich. Such a
family would have  $\bar \kappa = 3/2$.

\newsec{Conclusion}

The remarks of section five are built on a house of cards, namely,
on a chain of conjectures about some  relations between D-branes,
black holes, and number theory.
 The weakest
link by far in the chain of conjectures is the relation between
D-branes and arithmetic suggested in \arthatt.  Admittedly, it is
a long shot. Nevertheless, as we have shown, it would have
dramatic consequences if true. At worst, there are a couple of
interesting coincidences, so perhaps it
 deserves some closer scrutiny. It is curious that the
choice of arithmetic model has some relevance for ``predictions''
such as \rdeelim. Whether this turns out to be an interesting
feature or a fatal flaw of our discussion remains to be seen.

\bigskip
\centerline{\bf Acknowledgements}\nobreak
\bigskip

We would like to thank P. Candelas, E. Diaconescu, D. Goldfeld, D.
Rohrlich,
 P. Sarnak, J. Silverman, and S. Zhang
for helpful discussions and correspondence. We are particularly
grateful to B. Gross for important critical remarks on the draft.
GM  is supported by DOE grant DE-FG02-92ER40704. SM is supported
by an NSF postdoctoral fellowship and by a Yale Hellman
fellowship. GM also thanks the Institute for Advanced Study for
hospitality and the Monell foundation for support during the
completion of this work; SM is similarly indebted to Harvard
University for their hospitality.

\bigskip
\bigskip
\bigskip

\appendix{A}{:~Large Values of $\f{L'}{L}(1,-p)$, by
  Stephen D. Miller}

\bigskip

In the above paper    we made use of the fact that the generalized
Riemann hypothesis (GRH) implies that $$\lambda(D) =
O(\log\log|D|),$$ where $$\lambda(D)={L'(1,D) \over L(1,D)}.$$
Also, we discussed the implications of how the $\log\log|D|$ rate
is in fact optimal. The upper bound is well-known; a proof can be
found in \stark, for example.  The purpose of this appendix is to
prove the lower bound:

\thm{A(=4.1 above)} Let $\lambda(D)={L'(1,D) \over L(1,D)}$ and
$D=-p$. As $p\rightarrow\infty$ among the primes which are 3
modulo 4,

$$\limsup\f{\lambda(D)}{\log\log|D|} \ge \f 12 $$ and
$$\liminf\f{\lambda(D)}{\log\log|D|} \le -\f 12.$$

The constant $\f 12$ is far from optimal; any non-zero constant
will do for the application in section 5 above.
 We remark that
the discriminants $D=-p$ are a very special kind; for one thing,
they are fundamental discriminants.  Therefore, the conclusion of
the theorem still holds if we weaken our conditions on $D$, e.g.
if it passes to $-\infty$ over all discriminants.

\lref\erdos{P. T. Bateman, S. Chowla, and P. Erd\" os, ``Remarks
on the size of $L(1,\chi)$," Publ. Math. Debrecen {\bf 1}, (1950),
165-182.}

\lref\elliot{P.D.T.A. Elliot, ``Probablistic Number Theory II:
Central Limit Theorems," Grund\-lehren der mathematischen
Wissenschaften {\bf 240}, Springer, New York, 1980.}

One can prove this theorem under the assumption of GRH by
modifying the technique of Littlewood (\littlewood), who was the
first to prove the analogous result for $L(1,D)$.
We will present a different proof here based on the methods of
\erdos~and \elliot, which has the advantage that it is
unconditional and that it allows us to infer the theorem over a
set as sparse as the primes.  Shorter proofs are possible, but we
present the one here because all of the background material is
contained in the standard reference \davenport.  Extremal theorems
of this type are proved in two steps: first one truncates the
dirichlet series definition for each individual $L(s,D)$, and then
averages the remaining finite sum over many $D$ to show the
theorem in the mean. Since one term must always meet the size of
the average, we conclude that there are large individual values.
We will use sieve methods for each of these steps.

\subsec{Character sums}

The material in this section is modified from \elliot, chapter 22.

\lem{1}[Lemma 22.4 of \elliot] Let $\chi_1,\ldots,\chi_J$ be
distinct primitive characters to moduli $\le Q$, and let $a_n$ be
arbitrary complex numbers.  Then $$\sum_{j=1}^J \left|
\sum_{n=1}^N a_n\chi_j(n)   \right|^2 \le (N+Jc_0Q\log
Q)\sum_{n=1}^N|a_n|^2$$ for some absolute constant $c_0>0$.

We will use Lemma 1 to make a series of estimates.  The exponents
we use are somewhat arbitrary and are hardly optimal.

 \prop{2} Let $Q$ be large and $(\log Q)^{20} \le U \le
Q^2,$ $b_p\in \IC~,~|b_p|\le 1$ for prime $p$.  Then at most
$O(Q^{3/4})$ of the distinct primitive characters $\chi$
 to moduli $\le Q$ violate
\eqn\proptwoeqn{\left|    \sum_{U<p< 2U} \f{b_p\log( p)
\chi(p)}{p} \right| \le U^{-1/10}.}

{\bf Proof:} Let $\chi_1,\ldots,\chi_J$ be these violating
characters. Let $a_n$ be defined by the expansion $$\sum_{n=1}^N
a_n\chi(n) = \left(\sum_{U<p< 2U}
\f{b_p\log(p)\chi(p)}{p}\right)^m~,~N=(2U)^m.$$ Then
$$a_{p_1\cdots p_m}=\f{b_{p_1}\cdots b_{p_m}}{p_1\cdots p_m}(\log
p_1)\cdots (\log p_m)\mu(p_1,\ldots,p_m) \le \f{m!}{U^m}(\log
2U)^m,$$ where $\mu(p_1,\ldots,p_m)\le m!$ is the multiplicity
from the expansion.
  Also, $$ \sum_{U<p <
2U}\left|\f{b_p\log p}{p}  \right|
\le
\sum_{U<p< 2U} \f{\log p}{p} = \log(2U/U)+o(1) \le 1$$ for large
$Q$.  Consequently $$\sum_{n=1}^N|a_n| \le \sum_{U< p_1,\ldots,p_m
< 2U} \f{|b_{p_1}|\cdots |b_{p_m}|(\log p_1)\cdots(\log
p_m)}{p_1\cdots p_m}\mu(p_1,\ldots,p_m) $$  $$= \left(
\sum_{U<p<2U}\left|\f{b_p \log p}{p} \right| \right)^m \le 1,$$and
so
 $$\sum_{n=1}^N |a_n|^2 \le (\max |a_n|) \sum_{n=1}^N |a_n|
\le \f{m!}{U^m}(\log 2U)^m.$$

By Lemma 1 applied to the exceptional characters
$\chi_1,\ldots,\chi_J$, $$J U^{-m/5}\le \left((2U)^m + Jc_oQ\log Q
\right)m!\left(\f{\log 2U}{U}\right)^m.$$  This estimate will be
more than sufficient for our purposes, and in fact we will use
$m!\le m^m$ to deduce the weaker
 \eqn\weaker{JU^{-m/5} \le (2m\log
2U )^m + Jc_0 Q\log Q\left(\f{m\log 2U}{U} \right)^m.}

Now we shall set $m=\left[\f{11}{8}\f{\log Q}{\log U} \right]+1,$
where $[\cdot]$ denotes the greatest-integer function.  In Lemma 3
we show that  $$U^{-m/5} > 2 c_0 Q\log Q\left(\f{m\log 2U}{U}
\right)^m$$ holds for large $Q$.  Granted this, \weaker\ now
implies $$JU^{-m/5} \le 2\cdot (2 m \log 2U)^m$$ for large $Q$. We
claim that $$J \le 2\left(2 m U^{1/5} \log 2U \right)^m
=O(Q^{3/4}).$$ If this were not so, then for any $C>0$ we would
have that

$$m\log(2m U^{1/5}\log 2U) \ge \f{3}{4} \log Q+C.$$ Adjusting $C$,
$$m\log 2 + \f{1}{5}m\log U + \log(m\log U) \ge \f{3}{4}\log Q+C$$
$$m\log 2 + \left(\f{11}{40}\log Q + \f{1}{5}\log U\right) +
\log\left( \f{11}{8}\log Q+\log U \right) \ge \f{3}{4} \log Q +C$$
$$m\log 2 + \f{27}{40}\log Q + \log(\f{27}{8}\log Q) \ge
\f{3}{4}\log Q + C.$$  Readjusting $C$ again slightly, we find
$$\f{11}{8}\f{\log Q}{\log U}\log 2 + \log\log Q \ge \f{3}{40}\log
Q + C$$ and finally $$\f{11}{8}\f{\log Q \log 2}{20 \log\log Q} +
\log\log Q \ge \f{3}{40}\log Q + C,$$ which is absurd.

{\bf $\spadesuit$}

\lem{3} Keeping the notation of the above proof, let $C$ be any
fixed constant. Then for $Q$ sufficiently large, $$U^{-m/5} > C
Q\log Q\left(\f{m\log 2U}{U} \right)^m.$$

{\bf Proof:}  Otherwise $$\f{4m}{5}\log U \le \log C + \log Q +
\log\log Q + m\log (m\log 2U),$$ which implies
$$\f{11}{8}\f{4}{5}\log Q \le (1+\delta) \log Q + m\log( m\log
2U)$$ for some $\delta>0$ and $Q$ large.  Continuing
$$(\f{1}{10}-\delta)\log Q \le m\log(m\log 2U)$$ $$\le
\left(\f{11}{8}\f{\log Q}{\log U} + 1 \right)\log\left(
\left(\f{11}{8}\f{\log Q}{\log U}+1\right)(\log U+2)\right)$$
$$\le \left(\f{11}{8}\f{\log Q}{\log U}+1   \right) \log\left(
\f{11}{8}\log Q + \log U + \f{11}{4}\f{\log Q}{\log U} + 2
\right).$$  Using $(\log Q)^{20} \le U \le Q^2$, introducing a
tiny $\delta'>0$, and taking $Q$ larger yet, we get
$$(\f{1}{10}-\delta)\log Q \le \left(\f{11}{8}\f{(1+\delta')\log
Q}{20 \log\log Q} \right) \log\left(\f{11}{4\cdot 20}
\f{(1+\delta')\log Q}{\log\log Q}+\f{27}{8}\log Q + 2\right)$$
$$\le (1+\delta')\f{11}{160} \f{\log Q}{\log\log Q}\log(7/2 \log
Q).$$ This is a contradiction for large $Q$, because
$\f{1}{10}>\f{11}{160}$.

        {\bf    $\spadesuit$}

\prop{4} With the same notation as in Proposition 2 and Lemma 3,
but instead requiring that $U>Q^2$, at most $O(Q^{2/3}\log U)$
distinct primitive characters to moduli $\le Q$ have $$\left|
\sum_{U<p< 2U}\f{b_p\log(p)\chi(p)}{p}    \right| \ge Q^{-1/3}.$$

{\bf Proof:} Again, let $\chi_1,\ldots,\chi_J$ be these
characters.  By Lemma 1, $$JQ^{-2/3} \le (U+Jc_0Q\log Q)\sum_{U<p<
2U} \left|\f{b_p\log p}{p} \right|^2.$$  Since $|b_p|\le 1,$ the
prime number theorem guarantees that $$\sum_{U<p<2U}\left|
\f{b_p}{p} \log p \right|^2 \le \sum_{U<p<2U} \f{(\log
p)^2}{p^2}=O\left( \int_{U}^{2U} \f{(\log \xi)^2}{\xi^2}
\f{d\xi}{\log \xi} \right) = O\left(\f{\log U}{U}\right).$$
Therefore
 $$JQ^{-2/3}
\le C(U+Jc_0Q\log Q)\f{\log U}{U} = C_1 \log U + C_2 J \f{Q \log
Q}{U}\log U,$$ or $$J\left(1- C_2 U^{-1}Q^{5/3}\log Q \log U
\right)\le C_1 Q^{2/3} \log U.$$

Since $U>Q^2$, $$\f{Q^{5/3}}{U}\log Q \log U < 2 Q^{-1/3} (\log
Q)^2$$ becomes arbitrarily small for large $Q$, and we conclude
$$J=O(Q^{2/3}\log U).$$

{\bf    $\spadesuit$}

We will require a form of Dirichlet's theorem on the distribution
of primes in residue classes.  To fix notation we will set $li(x)$
to be the logarithmic integral $$li(x)=\int_2^\infty\f{dt}{\log
t}$$ and $$\pi(x,k,l)=\#\{p \le x | p~prime,p \equiv l~ (mod~k)
\}.$$
 See \davenport~for more details on this topic. Siegel and Walfisz proved

\thm{SW} Let $\epsilon>0$  be fixed.  Then there exists a constant
$c$ depending on $\epsilon$ such that
 $$\pi(x,k,l) =
\f{li(x)}{\phi(q)}+O(x \exp(-c\sqrt{x}))~~,~~(k,l)=1$$ for $x\ge
\exp{k^\epsilon}$.

Page proved a result which is also valid for smaller $x$, but with
a possible exception (related to the Landau-Siegel zero):

\thm{P}  Let $Q$ be a positive integer.  Then there exists a
constant $b>0$ such that for any modulus $k \le Q$ -- except for
perhaps one exceptional modulus $k_1$ and its multiples -- we have
$$\pi(x,k,l) = \f{li(x)}{\phi(q)} + O(x \exp(-b\sqrt{\log
x}))~~,~~(k,l)=1$$ for $x \ge \exp{(\log Q)^2} \Leftrightarrow Q
\le \exp (\sqrt{\log x}) $. The possible exception $k_1$ grows to
infinity with $Q$.

Recall the formula that $$\f{L'(1,\chi)}{L(1,\chi)} =
 -\sum_{p~prime} \f{\chi(p)\log p}{p} +
O(1).$$  The next proposition shows that one can truncate the sum.

\prop{5} For $Q$ large and $(\log Q)^{20} \le y \le Q^2$, all but
$O(Q^{11/12})$ distinct primitive characters $\chi$ to moduli $\le
Q$ have $$\f{L'(1,\chi)}{L(1,\chi)} = -\sum_{p \le y}
\f{\chi(p)\log p}{p} + O(1).$$

{\bf Proof:} Another form of the Siegel-Walfisz theorem gives us
that $$\f{L'(1,\chi)}{L(1,\chi)} = - \sum_{p \le \exp{Q^{1/8}}}
\f{\chi(p)\log p}{p} + O(1).$$ Given $y$ between $(\log Q)^{20}$
and $Q^2$, we shall dyadically decompose the range from
$(y,\exp{Q^{1/8}}]$ into intervals $(U_k,2U_k]$, $U_k=2^k y\le
\exp{Q^{1/8}}$.  Among these $O(Q^{1/8})$ ranges we will
distinguish those that come before and after the range containing
$Q^2$, designated as $(U_{k_0},2U_{k_0}]$.

First, for $k\le k_0$ $$\left| \sum_{y<p < 2^{k_0+1}y}
\f{\chi(p)\log(p)}{p}     \right| \le \sum_{k=0}^{k_0}
\left|\sum_{U_k < p < 2U_k} \f{\chi(p)\log(p)}{p} \right|$$ $$\le
\sum_{k=0}^{k_0}(2^k y)^{-1/10} =O(y^{-1/10}),$$ for characters
which satisfy the bound of Proposition 2 in each range. There are
at most $O(Q^{3/4}(k_0+1))=O(Q^{7/8})$ exceptions.

Secondly, for the $k>k_0$ $$\left|    \sum_{2^{k_0+1} < p \le
\exp{Q^{1/8}}}  \f{\chi(p)\log p}{p}    \right| \le Q^{-1/3}
\sum_{2^ky \le \exp{Q^{1/8}} }1$$ $$
 =O(Q^{-1/3+1/8}) = O(Q^{-5/24})=O(1)$$ with the exception of
$$O(Q^{2/3}\sum_{2^ky\le \exp{Q^{1/8}}} \log(2^{k+1} y))=
O(Q^{2/3}(Q^{1/8})^2)=O(Q^{11/12})$$ characters.

{\bf     $\spadesuit$    }

\subsec{Constructing a set of discriminants}

Having proven Proposition 5, we have now established that if $D$
is a fundamental discriminant, then $$\f{L'(1,D)}{L(1,D)} =
\sum_{p \le (\log|D|)^{20}} \left(\f{D}{p}\right)\f{\log p}{p} +
O(1)$$ except for $D$ in a very sparse set.  In fact, we may
usually invoke this conclusion for fundamental discrimants of the
form $D=-p$, $p$ a prime with $p \equiv~3~(4)$. We shall now
present a set of candidates for extreme values of $L(1,D)$.  While
we will not be able to directly show that they are large, we will
prove an average result which allows us to conclude that at least
one of them is.

{\bf The candidates:} ~We will pattern ourselves after the
argument in \erdos.  Let $x$ be a large parameter and set
$$y=\f{\sqrt{\log x}}{(\log\log x)^2}.$$ Enumerate the odd primes
$p_2,p_3,\ldots,p_m \le y$ and form the product $$M=8p_2p_3\cdots
p_m = 4 \exp(\sum_{p\le y} \log p)=\exp(y+o(y)).$$ Not all of the
moduli $$\f{M}{p_m},\ldots,\f{M}{p_1}$$ are divisible by the
exceptional modulis $k_1$ mentioned in Theorem P (applied to $x$).
Let $k=\f{M}{p_r}$ be the first of these listed above (i.e. with
$r$ the largest) which is not divisible by $k_1$; if none exists,
then set $r=m$.   The purpose of this construction is to guarantee
that $$\pi(x,k,l) = \f{li(x)}{\phi(k)} + O(x\exp(-b\sqrt{\log
x}))$$ for $l$ relatively prime to $k$.  Using the Chinese
remainder theorem, we can find two special residues $l_{\pm}~mod~
k$ such that

$$\left(\f{l_{\pm}}{p}\right)=\pm 1~~,~~p_r\neq p \le y. $$

Define the sets $$\CS_{\pm}=\{ -q~| \sqrt{x}\le q \le x,q ~prime,
q\equiv l_{\pm}\pmod{k},(q,k_1) =1\}-E_{\pm},$$ where $E_{\pm}$ is
the exceptional set of Proposition 5.  It follows that the
cardinality of $\CS_{\pm}$ is
$S_{\pm}=\f{li(x)}{\phi(k)}+O(x\exp{-b\sqrt{\log x}}))$. For any
$D\in \CS_{\pm}$, $$  \f{L'(1,D)}{L(1,D)} =\mp \sum_{p_r\neq p\le
y} \f{\log p}{p} - \sum_{y<p\le (\log
x)^{20}}\left(\f{D}{p}\right)\f{\log p}{p} + O(1)$$
\eqn\indiv{\f{L'(1,D)}{L(1,D)} = \mp \f12\log\log x +o(\log \log
x)- \sum_{y<p\le (\log x)^{20}}\left(\f{D}{p}\right)\f{\log p}{p}
+ O(1).}

In the next section, we will show that this sum is $o(1)$ on
average.

\subsec{Sifting and averaging}

Following \erdos, we will use a version of the R\'enyi sieve to
average \indiv .

\lem{R} Let $\CZ$ be a set of $Z=|\CZ|$ integers in a range of $N$
consecutive integers, $f(p) \le p$ and $Q(p) >1$
otherwise-arbitrary functions, and set $$\tau=\min_{p\le\sqrt{N}}
\f{f(p)}{p}~~,~~Q=\max_{p\le\sqrt{N}} Q(p).$$  Let $$Z(p,h)=\#\{
z\in \CZ \mid z \equiv ~h (p) \}.$$ Then for all but at most
$\f{2NQ^2}{Z\tau}$ ``abnormal'' primes $\le \sqrt{N}$ we have that
$$\left|  Z(p,h)-\f{Z}{p}      \right| < \f{Z}{pQ(p)}$$ holds
except for at most $f(p)$ ``irregular'' residues.

{\bf Proof of Lemma R}: This follows from Chebyshev's inequality
applied to the sieve inequality $$\sum_{p\le Q} \sum_{h=1}^p\left(
Z(p,h)-\f{Z}{p} \right)^2 \le (N+Q^2)Z,$$ (\davenport, p. 158).

{\bf   $\spadesuit$}

\lem{6}(\erdos) Keeping the same notation as Lemma R, assume $p$
is a normal prime.  Then the number of $z\in\CZ$ in the irregular
residue classes $mod~p$ is $\le
Z\left(\f{f(p)}{p}+\f{1}{Q(p)}\right)$.

{\bf Proof of Lemma 6}: The number in regular classes exceeds
$$(p-f(p))\left( \f{Z}{p}-\f{Z}{p Q(p)}       \right) \ge Z-Z
\f{f(p)}{p} - \f{Z}{Q(p)}.$$

{\bf   $\spadesuit$ }

We will always take $f(p)=p/(\log p)^5$ and $Q(p)=(\log p)^5$.

The average of the sum in \indiv~is

\eqn\aved{\sum_{D\in\CS_{\pm}} \sum_{y<p\le (\log x)^{20} }
\left(\f{D}{p}\right)\f{\log p}{p}= \sum_{y<p\le (\log x)^{20}}
\sum_{h=1}^p \left(\f{h}{p}\right)\f{\log p}{p} \CS_{\pm}(p,h).}

\lem{7} This sum, restricted to the normal primes $p$, is $$
\sum_{y<p\le (\log x)^{20} normal} \sum_{h=1}^p
\left(\f{h}{p}\right)\f{\log p}{p} \CS_{\pm}(p,h)=o(S_\pm).$$

{\bf Proof of Lemma 7}: For normal $p$, $$ \sum_{h=1}^p
\left(\f{h}{p}\right) \CS_{\pm}(p,h) = \sum_{h=1}^p
\left(\f{h}{p}\right)\left( \CS_{\pm}(p,h)-\f{S_\pm}{p}\right)$$
$$\le \sum_{h~regular} \f{S_\pm}{pQ(p)} + \sum_{h~irreg.}
\CS_\pm(p,h)+\sum_{h~irreg.} \f{S_\pm}{p}$$ $$\le \f{S_\pm}{Q(p)}
+ S(p) \f{f(p)}{p} + \left(    \f{S_\pm}{Q(p)} + S(p) \f{f(p)}{p}
\right) \le \f{4 S_\pm}{(\log p)^5}.$$  The sum over the normal
primes $p$ is thus $$\le \sum_{p>y} \f{4 S_\pm}{p(\log p)^4}
=o(S_\pm).$$ {\bf $\spadesuit$}

\lem{8} The sum in \aved~over the abnormal primes is also
$o(S_\pm)$.

{\bf Proof of Lemma 8}: First, we demonstrate that there is at
most one abnormal prime $p\le \exp(y \log\log x)$.  This is
because $\CS_\pm(p,h) = \#\{D\in \CS_\pm \mid D \equiv h~(p)\}$.
By Theorem P applied to $x$ and the moduli $kp\le \exp(\sqrt{\log
x })$, there is at most one prime $p\le\exp(y \log\log x)$ which
violates $$\CS_\pm(p,h) = \f{S_\pm}{p-1} + O(x\exp(-b\sqrt{\log
x})) = \f{S_\pm}{p}+O(\f{S_\pm}{p^2})+ O(x\exp(-b\sqrt{\log
x})).$$  This means $p$ qualifies as a normal prime, because
$x\sim S_\pm \phi(k) \log x$.

By Lemma R, the number of abnormal primes $p\le \sqrt{x}$ is $\f{2
x (\log x)^{15}}{S_\pm} =O(\phi(k)(\log x)^{16})$.  We will bound
the sum over the abnormal primes $p$ individually as

$$\left|\sum_{D\in\CS_{\pm}} \sum_{y<p\le (\log x)^{20}~abnormal }
\left(\f{D}{p}\right)\f{\log p}{p}\right| \le S_\pm
 \sum_{y<p\le (\log x)^{20}~abnormal } \f{\log p}{p}$$
 $$\le S_\pm\left( \f{\log y}{y} +
 O\left(\phi(k)(\log x)^{16} \f{\log(y \log \log x)}{\exp(y \log \log x)}\right)
          \right)=o(S_\pm).$$

{\bf     $\spadesuit$}

{\bf Proof of Theorem A}: Lemmas 7 and 8 show that the average of
\indiv ~ over $\CS_{\pm}$ is  $$\f{1}{S_\pm}\sum_{D\in\CS_\pm}
\f{L'(1,D)}{L(1,D)} = \pm \f 12 \log \log x + o(\log \log x).$$ At
least one term must exceed this average, proving Theorem A. {\bf
$\spadesuit$}

 \listrefs

\bye